\documentclass{article}
\usepackage{arxiv}

\usepackage{lineno}
\usepackage[english]{babel}
\usepackage{color}
\usepackage{graphicx}
\usepackage{tabularx}
\usepackage{enumitem}
\usepackage[caption=false]{subfig} 
\usepackage{amsmath,amssymb,bm}
\usepackage[version=3]{mhchem}
\usepackage{verbatim}
\usepackage{multirow}
\usepackage{dcolumn}
\usepackage{mathtools}
\usepackage{float}
\usepackage{nicefrac}
\usepackage{siunitx}

\usepackage[utf8]{inputenc}
\usepackage[T1]{fontenc}
\usepackage{hyperref}
\usepackage{url}
\usepackage{booktabs}
\usepackage{amsfonts}
\usepackage[nopatch=footnote]{microtype}
\usepackage{lipsum}

\usepackage[numbers,sort&compress]{natbib}
\usepackage{doi}
\usepackage{setspace}
\usepackage{subfiles}
\usepackage{longtable}

\usepackage{xargs}
\usepackage[pdftex,dvipsnames]{xcolor}
\usepackage[colorinlistoftodos,prependcaption,textsize=tiny]{todonotes}
\newcommandx{\tr}[2][1=]{\todo[linecolor=red,backgroundcolor=red!25,bordercolor=red,#1]{#2}}
\newcommandx{\ft}[2][1=]{\todo[linecolor=Orange,backgroundcolor=Orange!25,bordercolor=Orange,#1]{#2}}

\usepackage{soul}
\usepackage[acronym,nopostdot,nonumberlist]{glossaries}

\newacronym{fmlip}{fMLIP}{foundational machine-learned interatomic potential}
\newacronym{mlip}{MLIP}{machine-learned interatomic potential}
\newacronym{mlips}{MLIPs}{machine-learned interatomic potentials}
\newacronym{pes}{PES}{potential energy surface}
\newacronym{aimd}{AIMD}{\textit{ab initio} molecular dynamics}
\newacronym{md}{MD}{molecular dynamics}
\newacronym{dft}{DFT}{density functional theory}
\newacronym{rdf}{RDF}{radial distribution function}
\newacronym{sse}{SSE}{solid-state electrolyte}
\newacronym{zbl}{ZBL}{Ziegler-Biersack-Littmark}
\newacronym{cg}{CG}{Clebsch-Gordan}
\newacronym{daf}{DAF}{Discovery Acceleration Factor}
\newacronym{ema}{EMA}{exponential moving average}

\makeatletter
\renewcommand{\@fnsymbol}[1]{\ifcase#1\or 1\or 2\or 3\or 4\or 5\or 6\or 7\or 8\or 9\or 10\else\@ctrerr\fi}
\makeatother

\title{
Equivariant Interatomic Potentials without Tensor Products
}

\author{ 
\href{https://orcid.org/0000-0002-9793-3450}{\includegraphics[scale=0.06]{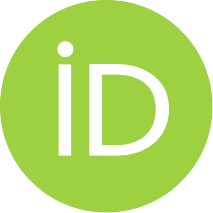}\hspace{1mm}Thiago~Reschützegger}\footnotemark[1]\\
IBM Research\\
Rio de Janeiro, 20031-170, RJ, Brazil\\
\texttt{thiagoreschutzegger@gmail.com}\\
\And 
\href{https://orcid.org/0000-0002-4293-3699}{\includegraphics[scale=0.06]{orcid.pdf}\hspace{1mm}Sarp Aykent}\\
Microsoft\\
Redmond, Washington, United States\\
\And
\href{https://orcid.org/0000-0002-5990-7497}{\includegraphics[scale=0.06]{orcid.pdf}\hspace{1mm}Gabriel Jacob Perin}\\
IBM Research\\
Rio de Janeiro, 20031-170, RJ, Brazil\\
\And
\href{https://orcid.org/0009-0007-3316-0608}{\includegraphics[scale=0.06]{orcid.pdf}\hspace{1mm}Bruno Henrique Nunes}\\
IBM Research\\
Rio de Janeiro, 20031-170, RJ, Brazil\\
\And
\href{https://orcid.org/0000-0002-5015-1443}{\includegraphics[scale=0.06]{orcid.pdf}\hspace{1mm}Flaviu Cipcigan}\\
IBM Research Europe\\
Daresbury, WA4 4AD, United Kingdom\\
\And
\href{https://orcid.org/0000-0003-4435-0507}{\includegraphics[scale=0.06]{orcid.pdf}\hspace{1mm}Rodrigo Neumann Barros Ferreira}\\
IBM Research\\
Rio de Janeiro, 20031-170, RJ, Brazil\\
\And
\href{https://orcid.org/0000-0003-1528-9292}{\includegraphics[scale=0.06]{orcid.pdf}\hspace{1mm}Mathias Steiner}\\
IBM Research\\
Rio de Janeiro, 20031-170, RJ, Brazil\\
\And
\href{https://orcid.org/0000-0003-2951-6740}{\includegraphics[scale=0.06]{orcid.pdf}\hspace{1mm}Fabian L.~Thiemann}\footnotemark[1], \footnotemark[2]\\
IBM Research Europe\\
Daresbury, WA4 4AD, United Kingdom\\
\texttt{fthiemann@microsoft.com}\\
}
\renewcommand{\undertitle}{}
\renewcommand{\headeright}{}

\begin{document}

\maketitle
\footnotetext[1]{To whom correspondence should be addressed.}
\footnotetext[2]{Present address: Microsoft Research AI for Science, Cambridge, CB1 2FB, United Kingdom.}

\begin{abstract}

Foundational machine-learned interatomic potentials have emerged as powerful tools for atomistic simulations, promising near first-principles accuracy across diverse chemical spaces at a fraction of the cost of quantum-mechanical calculations. However, the most accurate equivariant architectures rely on Clebsch-Gordan tensor products whose computational cost scales steeply with angular resolution, creating a trade-off between model expressiveness and inference speed that ultimately limits practical applications. Here we introduce Geodite, an equivariant message-passing architecture that replaces tensor products while incorporating physical priors to ensure smooth, well-behaved potential energy surfaces. Trained on the Materials Project trajectories dataset of inorganic crystals, Geodite-MP achieves accuracy competitive with leading methods on benchmarks for materials stability prediction, thermal conductivity, phonon-derived properties, and nanosecond-scale molecular dynamics, while running $3\text{--}5\times$ faster than models performing similarly. By combining predictive accuracy, computational efficiency, and physicality, Geodite enables faster large-scale atomistic simulations and high-throughput screening that would otherwise be computationally prohibitive.
\end{abstract}

\section*{Introduction}

The \gls{pes} governs atomic interactions and dynamics, thereby forming the foundation for computational predictions of material and molecular properties.
\Glspl{mlip} have emerged as powerful surrogates of the \gls{pes}, offering near first-principles accuracy at a fraction of the cost~\cite{Zuo-2020, Unke-2021, behler2007, Behler-2016, Thiemann-2024, Jacobs-2025, Friederich-2021, Deringer-2019, Chmiela-2018, vonLilienfeld-2020}. Their efficiency has enabled large-scale simulations of complex materials~\cite{Smith-2021, Musaelian-2023, Lu-2021, Fellman-2025, Nguyen-Cong-2021, Jia-2020, Deringer-2021}, protein dynamics~\cite{Wang-2024, Unke-2024, Kozinsky-2023}, and chemical reactions~\cite{Chen-2023, Zhang-2024, Wander-2025, Zhang-2024a, Guan-2023, Chen-2023a} extending accessible length and time scales far beyond those achievable with \gls{aimd}.

The increase in computing power~\cite{Pandey-2022, Stone-2007, Stone-2010}, architectural innovations~\cite{Schütt-2017, Gilmer-2017, Batzner-2022, Batatia-2022, Mazitov-2025}, and the size of available datasets~\cite{Barroso-Luque-2024, Chanussot-2021, Gharakhanyan-2025, Levine-2025, Sahoo-2025, Sriram-2024, Tran-2023, Deng-2023, Devereux-2020, Eastman-2023, Schmidt-2024, Huang-2025} have catalyzed a shift toward \glspl{fmlip}, which are trained on broad, chemically diverse datasets spanning large parts of the periodic table.
These general-purpose potentials have demonstrated strong transferability throughout chemical space~\cite{Batatia-2025, Wood-2025, Chen-2022, Merchant-2023, Zhang-2024b, Yang-2024} and can serve as a foundation for system-specific fine-tuning, significantly reducing the effort and cost of developing accurate specialized models~\cite{Radova-2025, Focassio-2025, Liu-2025, Pia-2025}.

The rapid progress of \glspl{fmlip} has driven the development of validation protocols and benchmarks to evaluate and compare models~\cite{Riebesell-2025, Chiang-2025, Póta-2025, Focassio-2025, Yu-2024}.
These span standard metrics such as energy and force errors, as well as high-order derivatives of the \gls{pes}, thermodynamic stability, geometry optimization, and \gls{md}~\cite{Fu-2025, Fu-2023, Miksch-2021, Stocker-2022, Liu-2023, Liu-2024, Riebesell-2025}. This diversity has helped identify systematic limitations in specific \gls{mlip} design choices such as non-conservative forces, restricted number of neighbors, insufficient short-range repulsion, and lack of smoothness which can result in structural instabilities, unphysical configurations, erroneous vibrational properties, discontinuities in binding curves, and energy drifts in microcanonical \gls{md} simulations~\cite{Witt-2023, Batatia-2025, Fu-2025, Rhodes-2025, Fu-2023, Morrow-2023, Deng-2025, Cărare-2025, Ranasinghe-2025}.

Beyond incorporating these physical priors in the model architecture to ensure reliable predictions, inference speed is critical in molecular simulations and high-throughput screening requiring hundreds of millions of model evaluations. A major bottleneck in many state-of-the-art equivariant architectures~\cite{Thomas-2018, Geiger-2022, Frank-2024, Batatia-2022} is the use of \gls{cg} tensor products to combine and transform steerable features while preserving SO(3) equivariance. Such tensor products enable more expressive and accurate models through higher angular information~\cite{Xie-2025, Batzner-2022, Liao-2023a, Liao-2023, Cen-2025}, but their $O(L_{\textrm{max}}^6)$ scaling constrains architectural choices. Beyond the standard considerations of cutoff radius and layer count, angular resolution ($L_{\textrm{max}}$) adds a third cost dimension, limiting the depth and scope achievable in practice despite the improved representational power.

Optimized CUDA kernels and specialized implementations have been explored to reduce this computational overhead in equivariant \glspl{mlip}~\cite{Geiger-2024, Tan-2025, Lee-2025}. Complementary efforts have focused on developing model architectures with a more favorable scaling with $L_\mathrm{max}$, for instance through sparsified tensor products that reduce the computational complexity to cubic scaling~\cite{Passaro-2023}. More recent approaches go a step further, showing that certain \gls{cg} operations are equivalent to inner products of high-degree steerable features, removing the cost of explicit tensor products while preserving equivariance and accuracy~\cite{Frank-2024, Frank-2022, Bharadwaj-2025, Aykent-2024, Zhang-2025}.

Among these, the Geometric Tensor Network (GotenNet) architecture, introduced recently by one of us~\cite{Aykent-2024}, achieves competitive or even superior accuracy to established \gls{mlip} architectures, such as MACE and Allegro~\cite{Batatia-2022, Batzner-2022}, on standard molecular benchmarks based on system-specific models~\cite{Chmiela-2022, Christensen-2020}. Notwithstanding these encouraging results, GotenNet has been evaluated primarily on energy and force errors, where physically motivated design considerations such as smoothness, short-range repulsion, or conservative forces are not strictly required and have limited impact on performance. Extending GotenNet into a robust \gls{fmlip} capable of performing reliable large-scale simulations, however, will require incorporating these fundamental constraints while maintaining computational efficiency and accuracy.

In this work, we present Geodite, an equivariant, tensor product-free architecture derived from GotenNet and enhanced with key physical priors and inductive biases. Based on this, we train a \gls{fmlip}, termed Geodite-MP, on the MPtrj dataset~\cite{Deng-2023}, a large-scale collection of relaxation trajectories from the Materials Project~\cite{Horton-2025} spanning nearly 160,000 diverse inorganic crystal structures. We systematically validate Geodite-MP across diverse tasks covering predictive accuracy and computational efficiency, comparing to other \glspl{fmlip} trained on the same dataset. Geodite-MP achieves competitive accuracy on established benchmarks such as Matbench Discovery~\cite{Riebesell-2025} and MDR~\cite{Loew-2025}, which probe a model's ability to identify a material's ground-state and reproduce its vibrational properties. Analysis of diatomic systems further shows that Geodite-MP produces smooth binding curves with correct short-range repulsion, whereas several other \glspl{mlip} exhibit notable deviations, including unphysical short-range attraction. To assess long-term performance in \gls{md}, we run 1~ns simulations of 49 \glspl{sse} at various temperatures, finding that Geodite-MP maintains stability and accurately reproduces the local structure observed in \gls{aimd} simulations~\cite{López-2024}. By avoiding computationally expensive \gls{cg} tensor products, Geodite-MP delivers this resolution while running approximately $5\times$ faster than Allegro-MP-L, $3\times$ faster than Eqnorm MPtrj, and $2.5\times$ faster than NequIP-MP-L, offering competitive accuracy at substantially lower computational cost. This makes Geodite-MP an ideal starting point for system-specific fine-tuning, large-scale simulations, and high-throughput screening.

\begin{figure}[ht!]
    \centering
    \includegraphics[width=1\linewidth]{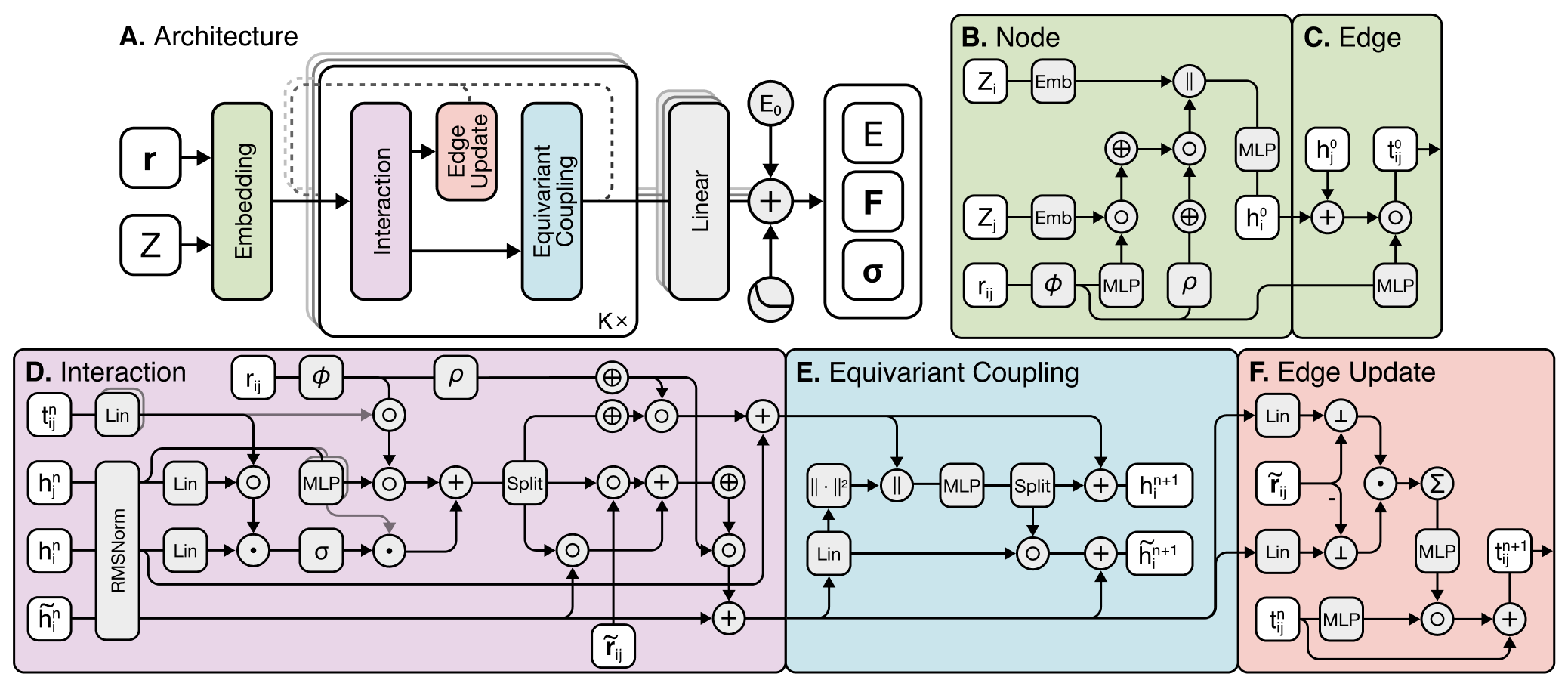}
    \caption{\textbf{Overview of the Geodite architecture.} \textbf{(A)} The system's potential energy $E$, atomic forces $\mathbf{F}$, and stresses $\bm{\sigma}$ are predicted by an equivariant message-passing neural network based on atomic positions $\mathbf{r}$ and types $\mathbf{Z}$.
    \textbf{(B--C)} Initial scalar node and edge features, $\textbf{h}^0_i$ and $\textbf{t}_{ij}^0$, are created based on atomic types and system's topology.
    \textbf{(D)} In an interaction layer, scalar and steerable node representations, $\textbf{h}^\mathrm{k}_i$ and $\tilde{\textbf{h}}^\mathrm{k}_i$, are refined based on neighbor information using modified self-attention and spatial filtering.
    \textbf{(E)} In the equivariant coupling block, information is exchanged between invariant and equivariant features to produce the output embeddings, $\textbf{h}^\mathrm{k+1}_i$ and $\tilde{\textbf{h}}^\mathrm{k+1}_i$, of the respective message passing layer.
    \textbf{(F)} Similarly, in the edge update, the revised steerable features and spherical harmonics representation of edge vectors, $\tilde{\textbf{r}}_{ij}$, are used to enhance the learnable edge representation, 
    $\textbf{t}_{ij}^{k+1}$. $\oplus$ denotes aggregation over neighbors, $\cdot\|\cdot$ denotes concatenation, $\bigcirc$ denotes element-wise product, $\|\cdot\|$ denotes vector norm, and $\phi_{\mathrm{eff}}$ and $\rho$ denote the effective radial basis embedding and learned atomic density factor, respectively.}
    \label{fig:architecture}
\end{figure}

\section*{Results}

\subsection*{Geodite Architecture}

We start by describing the Geodite architecture, which extends GotenNet~\cite{Aykent-2024} with physically motivated inductive biases and constraints. A comprehensive overview of the full architecture is provided in Fig.~\ref{fig:architecture}. Here, we first summarize the elements shared by GotenNet and Geodite at a high level, before describing the modifications introduced in Geodite to ensure smoothness, short-range repulsion, and correct asymptotic behavior. For a detailed description of the individual architectural components, the reader is referred to the original publication~\cite{Aykent-2024}. Geodite is an equivariant message-passing neural network that predicts energies $E$ from atomic positions $\mathbf{r}$ and species $Z$ (Fig.~\ref{fig:architecture}A), with forces $\mathbf{F}$ and stresses $\boldsymbol{\sigma}$ obtained via backpropagation following a conservative approach. The system is represented as a graph, with atoms as nodes and edges connecting atoms within a distance cutoff, $r_\mathrm{max}$. Initial latent features are produced by embedding layers illustrated in Fig.~\ref{fig:architecture}B and C, where each atom's type is encoded by a learnable embedding $\mathbf{Emb}(Z_i)$. These embeddings are aggregated with those of neighboring atoms, weighted by radial basis embedding modulated by a smooth cutoff envelope~\cite{Gasteiger-2019}, $\phi_{\mathrm{eff}}(r_{ij})$, and passed through a multilayer perceptron (MLP) to produce the initial invariant node features $\mathbf{h}_i^{(0)}$. In contrast to other \glspl{mlip}, Geodite constructs scalar edge features $\mathbf{t}_{ij}^{(0)}$ from the node features of connected atoms using element-wise multiplication with distance-dependent radial features to enhance the representation of interatomic interactions.

These initial node and edge features are then refined through a series of $K$ message-passing layers, each comprising three key operations shown in Fig.~\ref{fig:architecture}D–F. The interaction block (Fig.~\ref{fig:architecture}D) aggregates information from neighboring atoms to update scalar and steerable (i.e. higher order) node representations through a mechanism that combines modified self-attention and spatial filtering. Specifically, messages are constructed from scalar node features $\mathbf{h}^{\mathrm{k}}_j$ and edge features $\mathbf{t}^{\mathrm{k}}_{ij}$ and decomposed into three parts. The first forms the scalar messages, which are pooled over neighbors and used to residually update the central atom’s invariant representation $\mathbf{h}^{\mathrm{k}}_i$. The remaining components act as modulation coefficients that scale the neighbor’s steerable features $\tilde{\mathbf{h}}^{\mathrm{k}}_j$ and the steerable edge representations $\tilde{\mathbf{r}}_{ij}$, the latter being the spherical-harmonic projection of the normalized edge vectors. Combining these yields the equivariant messages, which are aggregated and applied as a residual update to the central atom’s steerable representation $\tilde{\mathbf{h}}^{\mathrm{k}}_i$. In the subsequent Equivariant Coupling block (Fig.~\ref{fig:architecture}E), these invariant and steerable node features interact to produce the updated representations ${\mathbf{h}}^\mathrm{k+1}_i$ and $\tilde{\mathbf{h}}^\mathrm{k+1}_i$. Finally, in the edge update (Fig.~\ref{fig:architecture}F), edge embeddings $\mathbf{t}_{ij}^{\mathrm{k}+1}$ are refined based on inner products between edge representations $\mathbf{r}_{ij}$ and node steerable features $\tilde{\mathbf{h}}^{\mathrm{k+1}}_i$ modified in the interaction block. Importantly, all operations preserve $O(3)$ equivariance while avoiding costly \gls{cg} tensor products through efficient inner-product computations.

Atomic energies are computed after multiple message-passing iterations through a multi-layer readout mechanism, which improves gradient flow and model expressiveness. The total system energy is obtained by summing atomic contributions, and forces are derived through automatic differentiation with respect to atomic positions. With the general framework established, we now describe the modifications introduced in Geodite to enforce physical priors, including smoothness and short-range repulsion.

\textbf{Residual energy and isolated atom limit.}

Geodite follows the standard approach~\cite{Batatia-2025a} of predicting residual atomic energies $E_i^{\mathrm{res}}$, which are added to the corresponding isolated-atom energies $E_i^{0}$ obtained from \gls{dft}. To ensure vanishing residual energy in the isolated-atom limit without constraining the model architecture, we introduce a vacuum embedding for each atomic species. The readout function is then applied only to the difference between atomic and vacuum embeddings, ensuring correct asymptotic behavior.

\textbf{Smoothness.} 
Certain architectural choices in GotenNet~\cite{Aykent-2024}, including softmax-based self-attention for edge weighting and biases in edge updates, can introduce discontinuities in the \gls{pes}. When neighbors cross the cutoff boundary, softmax renormalization produces abrupt changes in energy contributions, while learnable biases can offset and partially undo smooth cutoff envelopes, creating kinks in energies and spikes in forces. Geodite addresses these issues by replacing softmax with element-wise SiLU activations, as originally suggested in reference~\cite{Thölke-2021}, and removing biases from edge operations.
 
Additionally, in the equivariant coupling block, normalizing steerable features by their own magnitude can produce small discontinuities in forces and higher-order derivatives around perfectly symmetric configurations, where these features are naturally zero.
To prevent this and ensure smooth gradients, we normalize using the combined magnitude of steerable and invariant features.

\textbf{Short-range repulsion}. 
Accurate modeling of short-range repulsion is essential for simulation stability and physically correct binding curves, yet configurations in this regime are typically underrepresented in training data. Geodite explicitly models short-range repulsion using a term based on the functional form of the \gls{zbl} potential~\cite{Ziegler-1985}, similar to the implementation in ACEPotentials.jl~\cite{Witt-2023} and the MACE-MP-0b3 model~\cite{Batatia-2025}, but with learnable screening parameters that allow the model to refine the repulsion in regions where training data is available. To prevent the network from predicting unphysical energies at very short distances, we apply a smooth attenuation function to the interatomic distances, $r_{ij}$, 
\begin{equation}
\begin{aligned}
y(r_{ij}) &= \frac{a \, t^q}{1 + t^{q-p} + a \, t^q}, &
t &= \frac{r_{ij}}{r_{\rm cov}(Z_i) + r_{\rm cov}(Z_j)},
\end{aligned}
\end{equation}
 where $r_\mathrm{cov}$ is the covalent radius, $Z_i$ and $Z_j$ are atomic numbers, and $a = -2(p + q - 2qp)/(p^2 + p + q^2 + q)$ with learnable parameters $p$ and $q$ which are constrained to be positive. The effective radial distance embeddings, $\phi_{\mathrm{eff}}(r_{ij})$, are then obtained by multiplying the original radial basis functions, already modulated by a smooth envelope cutoff function, with the attenuation function $y(r_{ij})$. This ensures maximum gradient near the sum of covalent radii, providing high resolution in chemically relevant regions while smoothly suppressing network contributions as atoms approach unphysical overlap, allowing the \gls{zbl} term to dominate.

\textbf{Internal normalization.}

Message-passing architectures aggregate information from varying numbers of neighbors, which can lead to large variations in feature magnitudes across different coordination environments. To mitigate this and ensure stable, generalizable predictions, Geodite employs two complementary normalization strategies.

First, to address coordination-dependent scaling, we follow the approach introduced in MACE-MP~\cite{Batatia-2025}, and compute a smooth density factor for each edge:

\begin{equation}
\rho_{ij} = \tanh\left(\textrm{SiLU}\left(\mathbf{W}_\rho^\top \phi_\mathrm{eff}(r_{ij})\right)^2\right),
\end{equation}

where $\mathbf{W}_\rho$ are learnable weights and $\phi_\mathrm{eff}(r_{ij})$ denotes the effective radial basis embeddings evaluated at distance $r_{ij}$. Edge contributions are summed per atom to obtain a per-atom density factor $\rho_i$, which is shifted by 1 and used to normalize the aggregated invariant and steerable feature updates before residual addition.

The second normalization stabilizes internal representations across message-passing layers, preventing feature explosion or vanishing in deep models. We implement a RMS layer normalization similar to EquiformerV2~\cite{Liao-2023}, while using a unified statistic across all spherical harmonic degrees. This avoids discontinuities that can occur when individual degree-specific norms approach zero in symmetric environments, ensuring smooth energy surfaces and bounded feature magnitudes, analogous to the design choices in the Equivariant Coupling block.

\subsection*{Experiments}

We evaluate Geodite-MP across a variety of benchmarks spanning computational efficiency, predictive accuracy, physical behavior, and long-term \gls{md} stability, comparing its performance to other \glspl{fmlip} trained on the same MPtrj dataset. For completeness, we report energy and force errors on a subset of the sAlex dataset~\cite{Barroso-Luque-2024, Ghahremanpour-2018}, comprising systems not present in the MPtrj dataset, in Section B of the Supplementary Information.

\subsection*{Accuracy-Efficiency Trade-off on Matbench Discovery}

\begin{figure}[ht]
    \centering
    \includegraphics[width=1\linewidth]{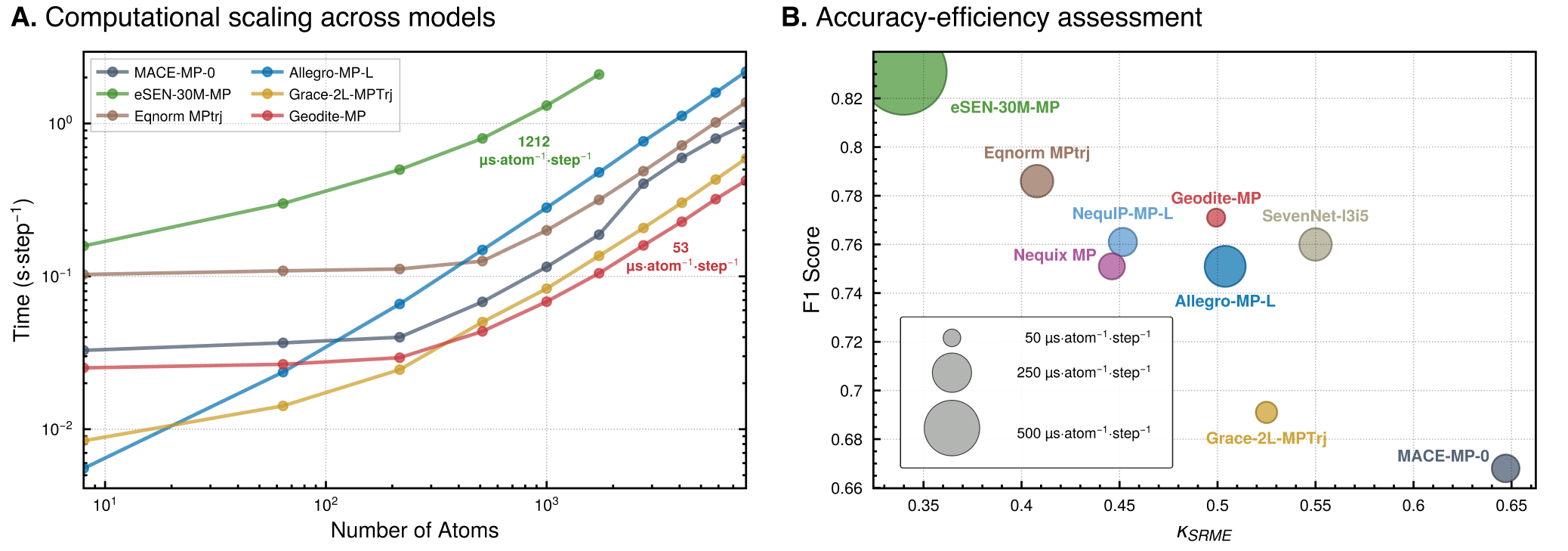}
    \caption{\textbf{Computational efficiency and Matbench Discovery accuracy trade-off across \glspl{fmlip}.} 
\textbf{A.} Inference time per molecular dynamics step as a function of system size. All models show increasing computational cost with system size, with different scaling behaviors emerging beyond 500 atoms. 
Beyond 500 atoms, all models scale linearly on a log-log plot, with vertical shifts reflecting differences in computational cost per atom.
\textbf{B.} Relationship between Matbench Discovery performance metrics (F1 score and $\kappa_{\textrm{SRME}}$) and computational efficiency. Marker size indicates inference speed, with smaller markers representing faster models. Models occupy different positions along the accuracy-efficiency trade-off, with some prioritizing predictive performance and others computational throughput.}
    \label{fig:speed_accuracy}
\end{figure}

We begin by evaluating computational efficiency across different \glspl{fmlip}. Fig.~\ref{fig:speed_accuracy}A shows the inference time per energy and force evaluation for different \glspl{fmlip} as a function of system size, measured during a short \gls{md} run. For systems larger than 500 atoms, Geodite-MP achieves the lowest inference times, converging to \SI{53}{\micro\second} per atom per step, corresponding to a 23$\times$ speed-up over eSEN-30M-MP~\cite{Fu-2025} (\SI{1212}{\micro\second} per atom per step). GRACE-2L-MPtrj~\cite{Bochkarev-2024} and Allegro-MP-L~\cite{Musaelian-2023} are competitive for smaller systems but scale less efficiently beyond this size.

We note that these timings were obtained using the ASE calculator interface. Since the efficiency advantages of Geodite arise from its model architecture rather than framework-level optimizations, these benefits are expected to be preserved across frameworks such as TorchSim~\cite{Cohen-2025}, which batches multiple independent systems on a single GPU, and the LAMMPS ML-IAP-Kokkos interface~\cite{NVIDIA-2025}, which enables multi-GPU parallelism in models with message-passing steps.

To put these results into context, the Matbench Discovery benchmark~\cite{Riebesell-2025} tests models on two key capabilities: predicting thermodynamic stability through convex hull distances and reproducing vibrational properties via higher-order energy derivatives. Performance is commonly summarized using F1 score for material classification and $\kappa_\mathrm{SRME}$ for thermal conductivity, which depends on accurate phonon spectra and higher-order PES derivatives~\cite{Póta-2025}. Fig.~\ref{fig:speed_accuracy}B combines these metrics with computational efficiency for a selection of models, plotting F1 score versus $\kappa_\mathrm{SRME}$ and scaling markers by per-step inference time. Geodite-MP achieves a high F1 score, outperformed only by Eqnorm MPtrj and eSEN-30M-MP, and $\kappa_\mathrm{SRME}$ of 0.499, comparable to Allegro MP-L and only slightly below the top four models. This competitive performance stems from using a 6 Å cutoff, four interaction layers, and $L_\mathrm{max}=2$, which allow for an expressive representation of atomic environments. By avoiding computationally expensive CG tensor products, Geodite-MP delivers this resolution at a fraction of the cost of the other models, offering a favorable efficiency–accuracy balance compared with previous architectures.

To complement Matbench Discovery, Table \ref{tab:matbench_discovery} presents a full overview of all metrics along with model size and cutoff. The \gls{daf} quantifies how effectively a model identifies stable structures relative to random selection. MLIPs using non-conservative forces, such as eqV2 S DeNS~\cite{Liao-2023a}, can achieve very high F1 and DAF values but at the expense of $\kappa_\mathrm{SRME}$, reflecting systematic errors in the \gls{pes} curvature. Geodite-MP, conversely, provides a balance across all metrics, often matching or exceeding established architectures such as NequIP-MP-L or Allegro-MP-L while using $3\times$ and $5\times$ fewer parameters, respectively. This includes regression metrics, with an energy MAE of 0.041~eV/atom and $R^2 = 0.806$, and accurate prediction of the relaxed ground-state geometries, measured by the root mean square displacement (RMSD) between predicted and reference structures. These results demonstrate Geodite-MP's capability to capture both energetic and structural aspects of the PES.

\begin{table}[H]
    \caption{\textbf{Matbench Discovery benchmark for foundation MLIPs.}
Performance metrics for foundational models trained on MPtrj dataset. F1 score measures ability to identify thermodynamically stable materials, while $\kappa_{\textrm{SRME}}$ quantifies accuracy in predicting thermal conductivity through phonon calculations. DAF indicates discovery acceleration relative to random selection. Energy MAE and force regression R$^2$ assess prediction quality. Cutoff radius $r_{\mathrm{max}}$ and parameter count reported for each architecture. $(\uparrow / \downarrow)$ indicate whether higher or lower values are preferred.}
    \label{tab:matbench_discovery}
\resizebox{\textwidth}{!}{
        \begin{tabular}{lrc |ccccccccc}
             \toprule
             \midrule
             {Model} & 
              \#Params (M)& 
             $r_{\mathrm{max}}$ & {CPS}$\uparrow$ & 
             Accuracy $\uparrow$ 
             &F1 $\uparrow$ &  DAF $\uparrow$  & Prec $\uparrow$ & MAE $\downarrow$ & R$^2$ $\uparrow$ & $\kappa_{\mathrm{SRME}}$ $\downarrow$ & RMSD $\downarrow$ 
             \\ 
             \midrule
             \midrule
            eSEN-30M-MP  &  30.1 & 6& 0.797 & 0.946 & 0.831 & 5.260 & 0.804& 0.033 & 0.822& 0.340& 0.075 \\
            Eqnorm MPtrj&  1.31 & 6& 0.756 & 0.929 & 0.786 & 4.844 & 0.741& 0.040 & 0.799& 0.408& 0.084\\
            Nequix-MP-L &  0.708 & 6& 0.729 & 0.914 & 0.751 & 4.455 & 0.681&  0.044& 0.782& 0.446& 0.085\\
            NequIP-MP-L &  9.6 & 6& 0.733 & 0.921 & 0.761 & 4.704 & 0.719& 0.043 & 0.791& 0.452& 0.086 \\
            Allegro-MP-L & 5.00 & 6& 0.720 & 0.915 & 0.751 & 4.516 & 0.690 & 0.044 & 0.778& 0.504& 0.082 \\
            DPA-3.1-MPtrj &  4.81 & 6& 0.718 & 0.936 & 0.803 & 5.024 & 0.768& 0.037 & 0.812& 0.650& 0.080\\
            SevenNet-l3i5 & 1.17 &  5& 0.714 & 0.920 & 0.760 & 4.629 & 0.708& 0.044 & 0.776 &  0.550 & 0.085 \\  
            MACE-MP-0 &  4.69& 6& 0.644 & 0.878 & 0.669 & 3.777 & 0.577& 0.057 & 0.697 &  0.647& 0.091 \\
            eqV2 S DeNS &  31.2& 12& 0.522 & 0.941 & 0.815 & 5.042 & 0.771& 0.036 & 0.788 &  1.676& 0.076 \\
            [0.3em]\hline\\[-0.8em]
            \textbf{Geodite-MP}   & 3.6 & 6  & 0.728 & 0.925 & 0.771 & 4.771 & 0.729 & 0.041 & 0.806 & 0.499 & 0.086 \\
         \midrule
            \bottomrule
        \end{tabular}
    }

\end{table}

\subsection*{MDR Benchmark}

Building on Matbench Discovery analysis, we next evaluate Geodite-MP on the MDR Phonon benchmark~\cite{Loew-2025}. While Matbench Discovery assesses vibrational properties via $\kappa_\mathrm{SRME}$ for roughly 100 materials, MDR broadens this evaluation to nearly 10,000 materials. Examining phonon-derived properties such as maximum phonon frequency $\omega_\mathrm{max}$, entropy $S$, free energy $F$, and heat capacity $C_v$ at constant volume, this benchmark primarily probes the smoothness of the PES and second-order derivatives beyond atomic forces.

Table \ref{tab:mdr} provides an overview of performance, measured as mean absolute error (MAE) for each metric, for Geodite-MP and a selection of established \glspl{fmlip}. Our model performs consistently across all metrics, with errors of 11 meV/atom for free energy, 32 meV/K/atom for entropy, 5 meV/K/atom for heat capacity, and 23 cm$^{-1}$ for maximum phonon frequency. This places it between the top-performing model, eSEN-30M-MP, and mid-range models such as SevenNet-l3i5. Notably, Geodite-MP achieves low errors across properties derived from both low-frequency acoustic and high-frequency optical modes.

\begin{table}[b]
    \caption{\textbf{MDR phonon benchmark for \glspl{fmlip}~\cite{Loew-2025}.}
Mean absolute errors for phonon properties computed from the finite displacement method. Evaluated on nearly 10,000 inorganic materials. Lower values indicate better agreement with \gls{dft} reference calculations. All models trained on MPtrj dataset.}
    \centering
    \resizebox{\textwidth*9/10}{!}{
        \begin{tabular}{lcccc}
            \toprule
            \midrule
            {Model} &
            \shortstack[c]{Free Energy MAE $\downarrow$\\ \footnotesize (meV/atom)} &
            \shortstack[c]{Entropy MAE $\downarrow$\\ \footnotesize (meV/K/atom)} &
            \shortstack[c]{Heat Capacity MAE $\downarrow$\\ \footnotesize (meV/K/atom)} &
            \shortstack[c]{Max Frequency MAE $\downarrow$\\ \footnotesize (cm$^{-1}$)} \\
            \midrule
            \midrule
            MACE-MP-0         & 24   & 60   & 13 & 61 \\
            GRACE-2L-MPtrj    & 9    & 25   & 5  & 40 \\
            SevenNet-0        & 19   & 48   & 9  & 40 \\
            SevenNet-l3i5     & 10   & 28   & 5  & 26 \\
            eSEN-30M-MP       & 5    & 13   & 4  & 21 \\
            NequIP-MP-L       & 7    & 20   & 4  & 20 \\
            Allegro-MP-L      & 11   & 30   & 8  & 55 \\
            \midrule                                      
            Geodite-MP        & 11   & 32   & 5  & 23 \\
            \midrule
            \bottomrule
        \end{tabular}
    }
    \label{tab:mdr}
\end{table}

\subsection*{Homonuclear Diatomics}

\begin{figure}[ht]
    \centering
    \includegraphics[width=1\linewidth]{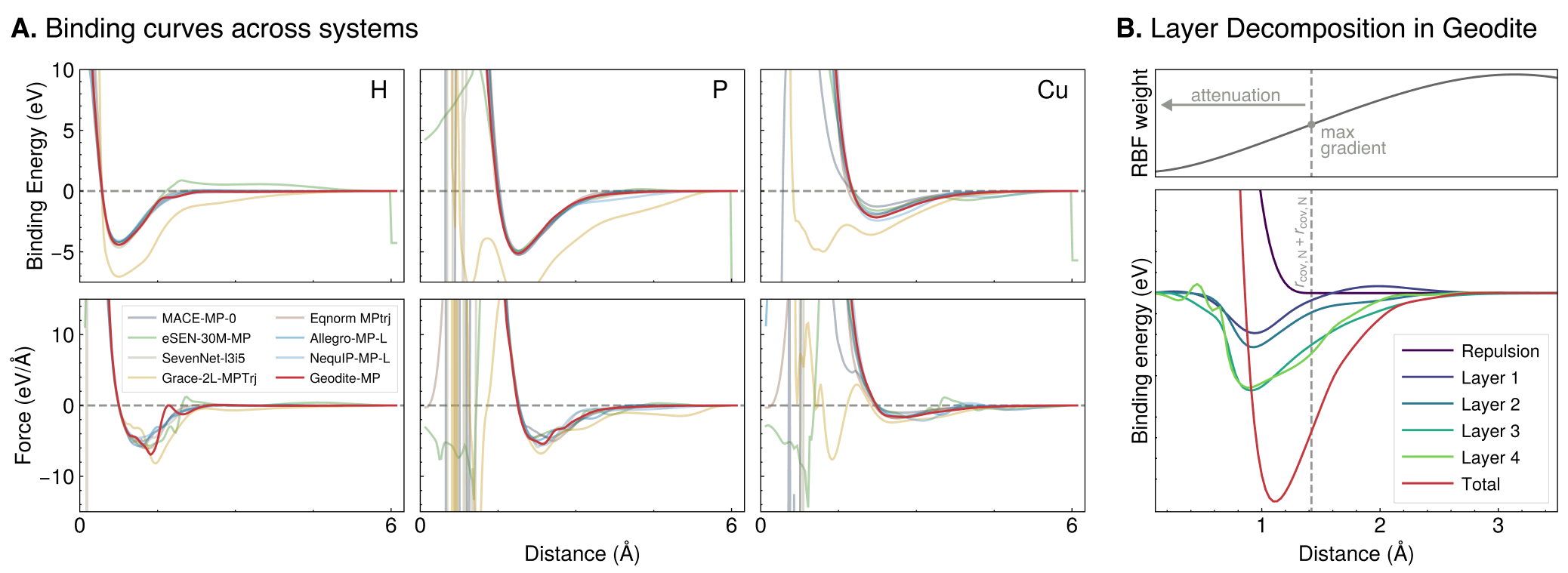}
    \caption{\textbf{Binding curves of homonuclear diatomics and asymptotic behavior.} 
\textbf{A.} Binding energy (top) and force (bottom) curves for homonuclear diatomics of H, P, and Cu as predicted by Geodite-MP and other \glspl{fmlip}.
\textbf{B.} Layer-wise decomposition of the binding energy surface for Geodite-MP demonstrates that the attenuation function (top) suppresses neural network contributions at short range, which allows the ZBL repulsive term to dominate as atoms approach unphysical overlap. Individual layer contributions decrease as the interatomic distance falls below the sum of covalent radii, while the repulsion term increases monotonically.}
    \label{fig:diatomics}
\end{figure}

\begin{table}[b]
    \caption{\textbf{Metrics for homonuclear diatomic binding curves.}
Metrics quantifying potential energy surface quality averaged over all element pairs in MPtrj dataset. Force and energy flips count sign changes in gradients. Tortuosity measures deviation from smooth monotonic behavior (1.0 indicates perfect smoothness). Jump metrics quantify cumulative discontinuities at inflection points. Conservation deviation measures mean absolute difference between forces and negative energy gradient. Spearman coefficients quantify the monotonicity of energy and force curves in the short-range repulsive regime, with values approaching $-$1 indicating physically correct repulsive interactions at close interatomic distances. $\downarrow$ indicates lower values are preferred. Metric definitions provided in Methods.}
    \centering
    \resizebox{\textwidth}{!}{
        \begin{tabular}{lrrrrrrrrrrrrr}
        \toprule
        \midrule
         & \shortstack{Force\\Flips\\$\downarrow$} & \shortstack{Force\\Total\\Variation\\$\downarrow$} & \shortstack{Force\\Jump\\$\downarrow$} & \shortstack{Energy\\Diff.\\Flips\\$\downarrow$} & \shortstack{Energy\\Grad\\Norm\\Max\\$\downarrow$} & \shortstack{Energy\\Jump\\$\downarrow$} & \shortstack{Energy\\Total\\Variation\\$\downarrow$} & \shortstack{Tortuosity\\$\downarrow$} & \shortstack{Conservation\\Deviation\\$\downarrow$} & \shortstack{Spearman\\Descending\\Force} & \shortstack{Spearman\\Ascending\\Force} & \shortstack{Spearman\\Repulsion\\Energy} & \shortstack{Spearman\\Attraction\\Energy} \\
        \midrule
        \midrule
        MACE-MP-0 & 2.83 & $2.3{\times}10^{4}$ & $3.0{\times}10^{3}$ & 2.62 & $1.6{\times}10^{4}$ & 21.3 & $1.3{\times}10^{3}$ & 2.54 & 39.9 & $-0.893$ & 0.096 & $-0.969$ & 0.633 \\
        eSEN-30M-MP & 4.34 & 104 & 46.6 & 3.93 & 25.1 & 0.287 & 20.2 & 1.86 & 0.127 & $-0.312$ & $-0.349$ & $-0.831$ & 0.789 \\
        MACE-MP-0b3 & 2.17 & 759 & 56.3 & 1.99 & 508 & 0.351 & 143 & 1.08 & 0.689 & $-0.909$ & 0.464 & $-0.967$ & 0.894 \\
        SevenNet-l3i5 & 5.02 & $4.5{\times}10^{12}$ & $5.3{\times}10^{10}$ & 4.31 & $3.1{\times}10^{12}$ & $1.7{\times}10^{6}$ & $1.4{\times}10^{11}$ & 2.31 & $9.5{\times}10^{9}$ & $-0.958$ & $-0.214$ & $-0.989$ & 0.545 \\
        GRACE-2L-MPtrj & 4.39 & $6.9{\times}10^{7}$ & 707 & 3.99 & $3.9{\times}10^{7}$ & 3.05 & $1.6{\times}10^{6}$ & 2.02 & $2.0{\times}10^{5}$ & $-0.906$ & 0.103 & $-0.858$ & 0.846 \\
        Allegro-MP-L & 1.56 & 207 & 27.6 & 1.39 & 145 & 0.047 & 103 & 1.04 & 0.139 & $-0.949$ & 0.710 & $-0.999$ & 0.907 \\
        Eqnorm MPtrj & 2.25 & 301 & 18.4 & 2.11 & 111 & 0.464 & 58.7 & 1.54 & 0.107 & $-0.852$ & 0.059 & $-0.963$ & 0.909 \\
        NequIP-MP-L & 1.93 & $3.5{\times}10^{3}$ & 3.30 & 1.86 & $3.1{\times}10^{3}$ & 0.305 & 726 & 1.01 & 2.21 & $-0.980$ & 0.526 & $-0.999$ & 0.903 \\
        \midrule
        Geodite-MP & 1.42 & $3.9{\times}10^{3}$ & 16.4 & 1.31 & $3.6{\times}10^{3}$ & 0.022 & 731 & 1.00 & 2.38 & $-0.989$ & 0.916 & $-0.999$ & 0.961 \\
        \midrule
        \bottomrule
        \end{tabular}
    }
    \label{tab:diatomics}
\end{table}

Following our phonon-based analysis of small displacements around equilibrium in many-body settings, we next probe the full interaction range using homonuclear diatomics across all chemical species encountered during training. This seemingly simple test is particularly revealing, as the MPtrj dataset contains no explicit diatomic configurations in the gas phase, requiring models to extrapolate correct two-body behavior from many-body training data.

Fig.~\ref{fig:diatomics}A shows binding curves for H, P, and Cu computed using a representative set of established \glspl{fmlip}. Corresponding binding curves from Geodite-MP for all elements are reported in Section A of the Supplementary Information. For the elements shown, Geodite-MP yields physically sensible binding behavior, equilibrium bond lengths consistent with other models, and exhibits smooth, monotonic short-range repulsion. In contrast, several other models show unphysical artifacts, including spurious oscillations, incorrect asymptotic behavior, and, most notably, attractive forces in the highly repulsive regime where nuclear-nuclear repulsion should dominate. These failures arise from a lack of explicit repulsive terms and unconstrained, arbitrary extrapolation in high-energy regions. Geodite-MP, conversely, describes short-range repulsion via the \gls{zbl} potential, while the attenuation function smoothly suppresses the learned potential in the highly repulsive regime. This ensures that the model focuses on the physically relevant interaction region, with \gls{zbl} dominating when atoms approach unphysical overlap, as illustrated in Fig.~\ref{fig:diatomics}B.

To quantify the smoothness of the binding curves, Table~\ref{tab:diatomics} reports metrics, originally introduced by MLIP Arena~\cite{Chiang-2025}, for all \glspl{fmlip}  averaged over all homonuclear pairs.
An overview of these metrics is included in the Methods section. Geodite-MP achieves the lowest values across most indicators, particularly excelling in force flips, energy difference flips, energy jumps, and tortuosity, the latter reflecting a near-perfect monotonic energy profile. Force jumps and maximum energy gradient norms remain moderate, showing that the model handles steep repulsive regions without discontinuities. In contrast, models such as SevenNet-l3i5 and GRACE-2L-MPtrj display extreme values in multiple metrics, with numerous force jumps and sign flips indicating non-monotonic behavior. Even among well-performing baselines, Geodite-MP consistently ranks at or near the top, demonstrating that its architectural design leads to a smoother \gls{pes} across diverse chemical systems.

\subsection*{Molecular Dynamics of Solid-State Electrolytes}

So far, our validation has been focused on static properties, providing insights into smoothness and accuracy of the learned \gls{pes}. However, these metrics offer limited information about a model’s ability to maintain stability in long-timescale \gls{md} simulations, where the iterative integration of the equations of motion can accumulate errors, driving the system away from training configurations potentially causing structural degradation or unphysical atomic configurations.

\begin{figure}[t]
    \centering
    \includegraphics[width=1\linewidth]{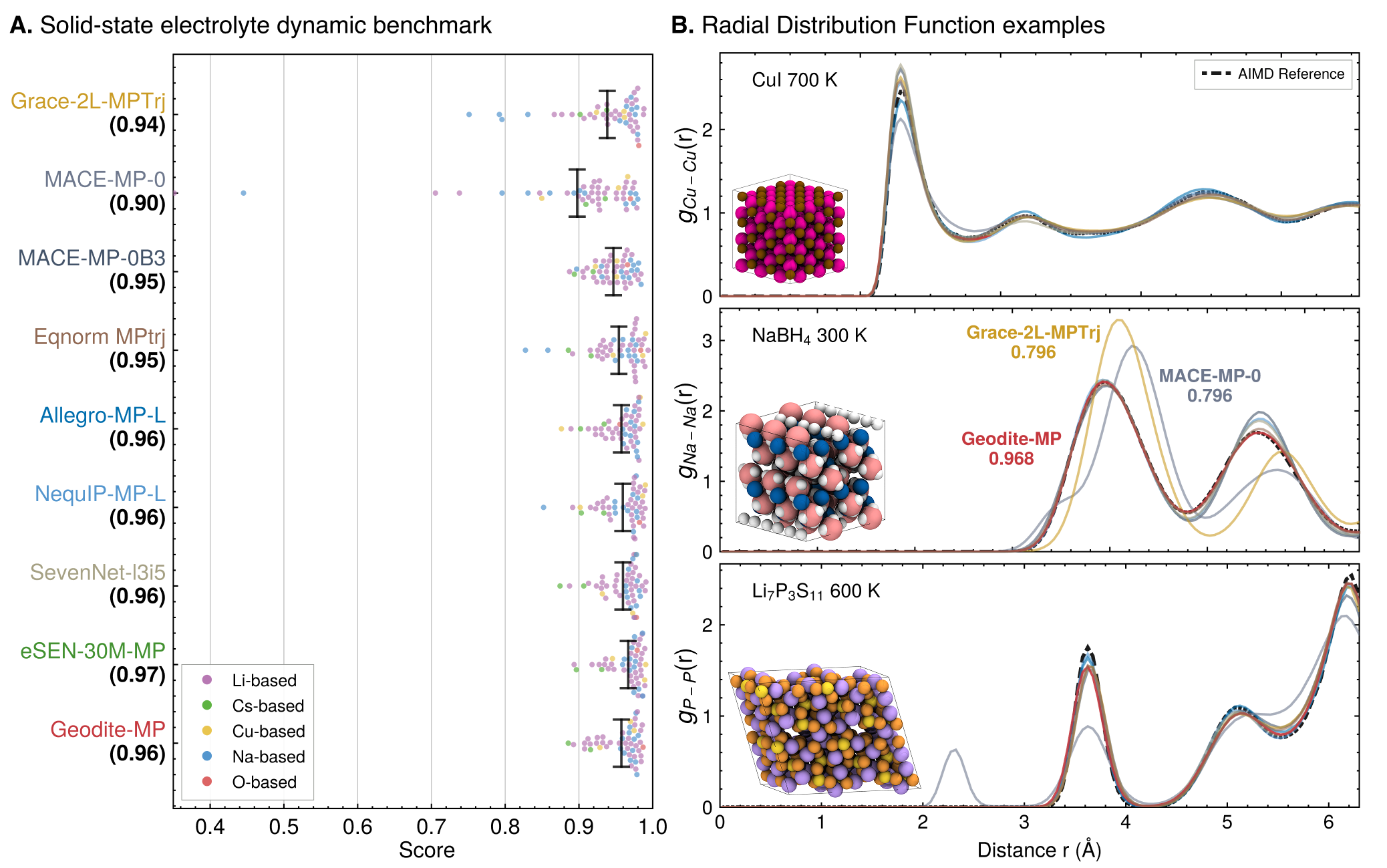}
    \caption{\textbf{Performance of \glspl{fmlip} on solid-state electrolyte molecular dynamics.} \textbf{A.} Benchmark scores aggregated across 49 solid-state electrolyte trajectories at various temperatures. Each dot represents a single system colored by the mobile ion species (Li, Cs, Cu, Na, or O). Scores quantify agreement between MLIP and \gls{aimd} \glspl{rdf}, with values closer to 1.0 indicating better reproduction. Models are ranked by mean score shown in parentheses. \textbf{B.} Representative \glspl{rdf} comparing MLIP predictions (colored lines) to \gls{aimd}  reference (black dashed) for three systems: CuI at 700~K (top), NaBH$_4$ at 300~K (middle), and Li$_7$P$_3$S$_{11}$ at 600~K (bottom). Insets show the crystal structure with mobile ions highlighted. Scores indicate quantitative agreement for the specific element pair shown.}
    \label{fig:rdfs}
\end{figure}

To address this, we extend our evaluation to 1 nanosecond (ns) \gls{md} simulations of 49 \gls{sse} systems at temperatures ranging from 300 to 1300~K (additional details in Section C of the Supplementary Information). Specifically, our analysis spans 16 different materials from the MPtrj dataset, comprising Li, Cs, Cu, Na, or O ions. \Glspl{sse} are highly relevant for battery applications and represent a challenging test case due to their diverse chemistries and coordination environments, where elevated temperatures may induce ion diffusion while the underlying sublattice remains stable~\cite{Famprikis-2019}. For each trajectory, we compare the atomic structure from each MLIP trajectory against reference \gls{aimd} simulations, originally performed by López, Rurali, and Cazorla, using pairwise \glspl{rdf} resolved for each element combination~\cite{López-2024}. Performance is subsequently quantified for each system via the minimum overlap score, explained in the Methods section, where a value of 1.0 indicates perfect agreement with the reference~\cite{Schran-2021}. After this manuscript was finalized, a complementary benchmark was introduced, focusing on Li–P–S electrolyte systems while extending the range of evaluated properties~\cite{Fragapane-2025}. Fig.~\ref{fig:rdfs}A presents the \gls{rdf} overlap scores for all 49 \gls{sse} systems across a selection of \glspl{fmlip}. Each point corresponds to a single trajectory and is colored by the ionic species. Overall, Geodite-MP ranks among the top performers with a mean score of 0.958, comparable to NequIP-MP-L (0.959), SevenNet-l3i5 (0.960), and eSEN-30M-MP (0.967). While these deviations should be considered minor given the comparatively short \gls{aimd} references, eSEN-30M-MP required several days to generate a 1~ns trajectory, whereas Geodite-MP completed the same simulation in under 5 hours on the same conditions.

Although the small numerical differences between top-performing models appear negligible, the qualitative trends across systems reveal how architectural choices influence \gls{md} performance. This is particularly evident when comparing MACE-MP-0 with its successor MACE-MP-0b3 with overlap scores of 0.898 and 0.947, respectively. This is likely due to the inclusion of short-range repulsion, density-based normalization, and an attenuation function, features we also incorporated into the Geodite architecture. Fig.~\ref{fig:rdfs}B further illustrates these trends via selected pairwise \glspl{rdf} for three \glspl{sse}. For CuI at 700 K, all models closely reproduce the AIMD reference for the Cu-Cu RDF, accurately capturing the liquid-like structure imposed by copper ion diffusion. Greater variations are observed for NaBH$_4$, where MACE-MP-0 and GRACE-2L-MPtrj predict higher peaks in the first coordination shell, with the latter indicating a more structured and less mobile sodium sublattice. For Li$_7$P$_3$S$_{11}$, only MACE-MP-0 predicts melting, highlighted by the emerging P-P RDF peak around 2.3~Å. Across all systems, Geodite reproduces the AIMD reference structure with high fidelity, demonstrating consistent reliability across diverse materials.

\section*{Discussion}

We introduced Geodite, a tensor product-free equivariant message-passing architecture for \glspl{mlip}. By replacing explicit \gls{cg} tensor products by inner products of steerable features, Geodite preserves full rotational equivariance while substantially reducing computational cost. This architectural choice enables deeper networks, larger cutoffs, and higher angular resolution within practical computational budgets, positioning Geodite as a scalable alternative to tensor product-based \glspl{mlip}. By incorporating smoothness constraints and physical priors, including short-range repulsion, directly into the architecture, Geodite ensures a physically consistent and smooth \gls{pes} suitable for atomistic simulations.

To demonstrate the practical capabilities of our architecture, we trained Geodite-MP on the Materials Project trajectory (MPtrj) dataset. Despite using $L_{\mathrm{max}}=2$ together with a cutoff radius of 6 Å and four message-passing layers, a setup that is computationally prohibitive for many tensor product-based architectures such as MACE,  Geodite-MP achieves state-of-the-art inference speed within the ASE framework. In particular, Geodite-MP is more than 2× and 4× faster than the established \glspl{fmlip} MACE-MP-0b3 and Allegro-MP-L, respectively, in terms of per-atom, per-step inference time. We evaluated Geodite-MP across benchmarks probing the smoothness, accuracy, and higher-order derivatives of the \gls{pes}, spanning tasks from materials discovery to phonon and thermal conductivity predictions, diatomic binding curves, and long-timescale \gls{md} simulations. Across all tasks, it demonstrates consistently competitive performance, ranking among the top models.

These results indicate the large potential introduced by the Geodite infrastructure and lay out promising avenues for further development. While Geodite-MP was trained exclusively on the MPtrj dataset, incorporating larger or more diverse datasets could improve transferability. In particular, training on recently released large-scale datasets such as OMAT24~\cite{Barroso-Luque-2024} for inorganic materials or OMOL25~\cite{Levine-2025} for organic molecules could further enhance accuracy and extend Geodite's applicability to broader chemical domains. The current architecture uses a local interaction cutoff and does not account for long-range interactions, which could extend applicability to polar and charged systems. Moreover, further optimization on modern hardware could further reduce computational cost and improve scalability.

Despite these limitations, the combination of accuracy, smoothness, and efficiency demonstrated by Geodite-MP highlights its practical potential for large-scale atomistic modeling. 
In practice, this efficiency at comparable accuracy enables large-scale, long-timescale simulations, high-throughput screening, and exploration of chemical spaces that would be significantly more expensive and potentially prohibitive with slower tensor product-based \glspl{mlip}.

\section*{Methods}

\subsection*{Model training and hyperparameters}

\textbf{Training data and preprocessing.} We trained Geodite-MP on the MPtrj dataset~\cite{Deng-2023}, a large-scale collection of \gls{dft} relaxation trajectories from the Materials Project~\cite{Horton-2025}. The dataset comprises approximately 1.6 million structures spanning nearly 160,000 distinct inorganic materials, calculated using the PBE exchange-correlation functional with spin polarization and Hubbard U correction. We randomly partitioned the data into 90\% for training and 10\% for validation, using a fixed seed for reproducibility.

\textbf{Model architecture and hyperparameters.} The Geodite-MP model uses 192 hidden channels, 4 message-passing layers, and a maximum angular momentum of $L_{\mathrm{max}}=2$. Radial interactions are encoded using 12 non-trainable Bessel basis functions. The model employs 8 attention heads, a dropout rate of 0.2, and SiLU activation functions for differentiability.

\textbf{Loss function.} We trained the model using a weighted sum of Huber losses for energy, forces, and stress:
\begin{equation}
\mathcal{L}  (\hat{E}, E,  \hat{F}, F, \hat{\bm{\sigma}},\bm{\sigma}) = \lambda_E \mathcal{L}_{\mathrm{Huber}}\left(\frac{\hat{E}}{N \sigma_E}, \frac{E}{N \sigma_E}\right) + \lambda_F \mathcal{L}_{\mathrm{Huber}}\left(\frac{\hat{\mathbf{F}}}{\sigma_F}, \frac{\mathbf{F}}{\sigma_F}\right) + \lambda_{\sigma} \mathcal{L}_{\mathrm{Huber}}\left(\frac{\hat{\bm{\sigma}}}{\sigma_{\sigma}}, \frac{\bm{\sigma}}{\sigma_{\sigma}}\right),
\end{equation}
where $\hat{E}$, $\hat{\mathbf{F}}$, and $\hat{\bm{\sigma}}$ denote predicted energy, forces, and stress, respectively, while $E$, $\mathbf{F}$, and $\bm{\sigma}$ are the corresponding \gls{dft} reference values and $\lambda_E, \lambda_F, \lambda_{\sigma}$ are hyperparameters. The energy is normalized by the number of atoms $N$, and all predictions and targets are divided by the standard deviation of each property in the training set ($\sigma_E$, $\sigma_F$, $\sigma_{\sigma}$) to balance the contributions of different quantities. The Huber loss with delta $\delta = 0.01$ was used for all three terms to reduce sensitivity to outliers.

\textbf{Optimization and training protocol.} Training was performed in two stages using the AdamW optimizer~\cite{Kingma-2017} with weight decay~\cite{Loshchilov-2019} of 0.001 and betas of (0.90, 0.999). In the first stage, we trained for 50 epochs with a batch size of 6 structures per GPU across 16 GPUs, using equal weights $\lambda_E = \lambda_F = \lambda_{\sigma} = 1$. In the second stage, we increased the number of GPUs to 64 and trained for 200 epochs, effectively quadrupling the batch size while increasing the force weight to $\lambda_F = 5$ to improve force prediction accuracy and PES smoothness. This two-stage approach resulted in approximately the same number of optimization steps per stage. During the first stage, the learning rate followed a warmup-linear-decay schedule, starting from $5 \times 10^{-6}$, warming up to $4.5 \times 10^{-4}$ over 2000 steps, then linearly decaying to $5 \times 10^{-6}$ over 800,000 steps. We applied exponential moving average (EMA) of model weights with a decay rate of 0.995, updating it at every training step. Gradient clipping was performed using the z-score method with $\alpha = 0.97$ and threshold $z_{\mathrm{thresh}} = 2.5$ ~\cite{Kumar-2025}.

\subsection*{Computational efficiency benchmark}

We measured wall-clock time per \gls{md} step for cubic silicon using supercells of increasing size, keeping density and average number of neighbors constant by adjusting the lattice parameters. All models ran through their respective ASE calculators ~\cite{HjorthLarsen-2017} on a single NVIDIA A100 GPU. For each step, the graph was rebuilt, stressing both neighbor-list generation and network inference for all models. Each model used its most competitive publicly available settings, such as enabling cuEquivariance operations, setting Torch compilation to on, and, for Allegro and NequIP, using Ahead-of-Time Inductor (AOTI) compilation. System sizes ranged from 8 to 8000 atoms. Inference cost was measured as the time per step divided by the number of atoms in the system, averaged over multiple steps to ensure stable timing measurements.

\subsection*{Matbench discovery}

Following the benchmark protocol, we performed structure relaxations using the FIRE optimizer with a force convergence criterion of $f_{\mathrm{max}} < 0.01$~eV/\AA\ and a maximum of 500 steps. We computed energies for each relaxed structure, and stability was determined by comparison to known ground states in the Materials Project database. For thermal conductivity predictions, we computed phonon properties using the finite displacement approach with 0.01~\AA\ perturbations after geometry optimization. We refer the reader to the Matbench Discovery website for detailed definitions of all metrics.

\subsection*{MDR}

Following the approach of Loew et al., we first optimized each structure using the FIRE algorithm with a force tolerance of $f_{\mathrm{max}} = 0.01$~eV/\AA\ and a maximum of 500 steps. We then calculated phonon properties using finite differences with 0.01~\AA\ atomic displacements in the harmonic approximation. From the phonon density of states, we computed the maximum phonon frequency $\omega_{\mathrm{max}}$, vibrational entropy $S$, Helmholtz free energy $F$, and constant-volume heat capacity $C_v$ at 300~K.

\subsection*{Diatomics}

We followed the protocols established by the MLIP Arena~\cite{Chiang-2025} to evaluate binding curves for homonuclear diatomics. For each element, we computed total energy as a function of interatomic distance by sampling 200 points spaced from $0.01$~\AA\ to $6.1$~\AA. Forces are obtained from automatic differentiation of the energy with respect to atomic positions. To quantify smoothness and physical behavior, we computed metrics following the MLIP Arena implementation, evaluated within element-specific distance ranges $[r_{\mathrm{min}}, r_{\mathrm{max}}]$, where $r_{\mathrm{min}} = 0.5 r_{\mathrm{cov}}$ and $r_{\mathrm{max}} = 6.0$~\AA, with $r_{\mathrm{cov}}$ denoting the covalent radius. 

Force flips count sign changes in force after rounding forces below $0.01$~eV/\AA\ to zero. Energy difference flips similarly count sign changes in consecutive energy differences. Tortuosity quantifies the deviation from the smooth repulsion-attraction behavior:

\begin{equation}
\tau = \frac{\sum_i |E(r_i) - E(r_{i+1})|}{|E(r_{\mathrm{min}}) - E_{\mathrm{min}}| + |E_{\mathrm{min}} - E(r_{\mathrm{max}})|},
\end{equation}

where $E_{\mathrm{min}}$ is the minimum energy (at equilibrium distance $r_{\mathrm{eq}}$), and values near $1.0$ indicate smooth energy profiles. Force and energy jumps (denoted as $E_j$) quantify cumulative discontinuities at flip points:

\begin{equation}
\begin{split}
E_j = \sum_{r_i \in [r_{\mathrm{min}}, r_{\mathrm{max}}]} \left|\text{sign}\left[E(r_{i+1}) - E(r_i)\right] - \text{sign}\left[E(r_i) - E(r_{i-1})\right]\right| \\
\times \left(\left|E(r_{i+1}) - E(r_i)\right| + \left|E(r_i) - E(r_{i-1})\right|\right),
\end{split}
\end{equation}
with force jump computed analogously. Conservation deviation measures the mean absolute difference between forces and the negative energy gradient:

\begin{equation}
\text{Conservation deviation} = \langle |F + \partial E/\partial r| \rangle.
\end{equation}

Spearman rank correlations assess monotonicity separately for repulsive ($r < r_{\mathrm{min}}$) and attractive ($r > r_{\mathrm{min}}$) regions, where $r_{\mathrm{min}}$ denotes the distance at which energy or force reaches its minimum value.

\subsection*{Local structure benchmark for solid-state electrolytes} \label{methods-sse}

To evaluate the performance of \glspl{mlip} in conducting long \gls{md} simulations under realistic conditions,  we focus on the group of \glspl{sse}, which represent a suitable test case due to their diverse chemical environments, ionic transport behavior, and relevance for battery applications. As a reference, we use the extensive dataset ~\cite{López-2024} of \gls{aimd} simulations produced by López, Rurali, and Cazorla, covering multiple chemical families of \glspl{sse}. While both our training set from the Materials Project~\cite{Horton-2025} and the reference \gls{aimd}  simulations use the PBE functional~\cite{Perdew-1996}, only the former calculations were spin-polarized and include a Hubbard U correction for transition metals in the presence of oxygen or fluorine. To enable a fair comparison, we therefore restrict our study to pristine, stoichiometric \glspl{sse} free of transition metals, and additionally exclude systems that melt in \gls{aimd} or have reference trajectories shorter than 30~ps. In total, this results in 16 different \glspl{sse} comprising Li, Na, Cu, Cs, or O ions, simulated at temperatures ranging from 300 to 1300~K, yielding 49 reference trajectories. An overview of all systems, including their chemical composition, system size, and simulation temperature, is provided in Section C of the Supplementary Information.

The \gls{mlip}-based simulations are carried out in the canonical (NVT) ensemble
using a timestep of 1.5~fs identical to the reference \gls{aimd}  simulations. The temperature is controlled using a Nosé-Hoover chain thermostat ~\cite{Nosé-1984} with 10 chains and a recommended damping time of $100 \times$ the integration time step. For each system, we perform 700,000 integration steps, yielding a trajectory of 1.05~ns length. All \gls{md} simulations are performed using the ASE software package ~\cite{HjorthLarsen-2017}.

\gls{mlip} and \gls{aimd} trajectories are compared based on their local atomic structure, characterised by the pairwise \gls{rdf}, $g(r)$, resolved for all combinations of chemical elements. The \glspl{rdf} are computed over the entire trajectory, sampling every 10th frame (i.e., every 15 fs), after discarding the initial 5~ps and 30~ps of the \gls{mlip} and \gls{aimd} trajectories, respectively, to allow for equilibration. We use a consistent bin width of $0.05$~\AA, with maximum radial distance set to half of the shortest periodic image separation, determined by the cell geometry of each individual system. A detailed overview of the \glspl{rdf} for both \gls{aimd} and Geodite-MP resolved by element for each system are shown in Section C of the Supplementary Information.

For each pair of chemical elements $i$ and $j$ in a system, we quantify the agreement between an individual \gls{mlip} and \gls{aimd} based on the overlap score
\begin{equation}
    d^{\mathrm{RDF}}_{i-j} = 1-\frac{\int_0^{r_{\mathrm{max}}}|g^{\mathrm{AIMD}}_{i-j}(r)-g^{\mathrm{MLIP}}_{i-j}(r)|~\mathrm{d}r}{\int_0^{r_{\mathrm{max}}}g^{\mathrm{AIMD}}_{i-j}(r) ~\mathrm{d}r + \int_0^{r_{\mathrm{max}}}g^{\mathrm{MLIP}}_{i-j}(r)~\mathrm{d}r} \mbox{ , }
\end{equation}
where $g_{i-j}(r)$ is the pairwise \gls{rdf} of elements $i$ and $j$ and $r_\mathrm{max}$ corresponds to the maximum resolved distance~\cite{Schran-2021}. Larger values of $d^{\mathrm{RDF}}_{i-j}$ indicate better agreement between model and reference, with a score of $1.0$ representing perfect overlap. We assign each system its minimum pairwise score, rather than the originally suggested average, to better differentiate between \glspl{mlip}.

\section*{Data Availability}

The code for the Geodite architecture implementation can be found at \url{https://github.com/IBM/materials/tree/main/models/pos_egnn/mlip}. The trained Geodite-MP model weights are available at \url{https://huggingface.co/ibm-research/materials.geodite}. Training and evaluation scripts are provided in the same repository. The code is released under the Apache 2.0 license.

This study was carried out using publicly available data from the Materials Project trajectories (MPtrj) dataset at \url{https://figshare.com/articles/dataset/Materials_Project_Trjectory_MPtrj_Dataset/23713842}~\cite{Deng-2023}. Reference solid-state electrolyte AIMD trajectories used for validation are available at \url{https://superionic.upc.edu/}~\cite{López-2024}.

\section*{Author Contributions}

\textbf{T.R.:} Conceptualization, Methodology, Software, Investigation, Formal analysis, Data curation, Writing - original draft.
\textbf{F.L.T.:} Conceptualization, Methodology, Software, Investigation, Formal analysis, Supervision, Data curation, Writing - original draft.
\textbf{S.A.:} Methodology, Formal analysis.
\textbf{G.J.P.:} Software, Formal analysis.
\textbf{B.H.N.:} Software, Formal analysis.
\textbf{F.C.:} Conceptualization, Software, Resources.
\textbf{R.N.B.F.:} Conceptualization, Resources.
\textbf{M.S.:} Conceptualization, Resources, Supervision.
All authors contributed to Writing - review \& editing.

\section*{Conflicts of Interest}

The authors declare no conflicts of interest.

\section*{Acknowledgements}
The authors thank Ilyes Batatia for helpful discussions and Cibrán López and Claudio Cazorla for sharing SSE data. We also thank Hendrik Kraß and Mohamad Moosavi for early testing of the model on out-of-distribution systems. This work was supported by the Hartree National Centre for Digital Innovation, a collaboration between the Science and Technology Facilities Council and IBM.

\clearpage

\bibliographystyle{unsrtnat}
\bibliography{references}

\begin{thebibliography}{109}
\providecommand{\natexlab}[1]{#1}
\providecommand{\url}[1]{\texttt{#1}}
\expandafter\ifx\csname urlstyle\endcsname\relax
  \providecommand{\doi}[1]{doi: #1}\else
  \providecommand{\doi}{doi: \begingroup \urlstyle{rm}\Url}\fi

\bibitem[Zuo et~al.(2020)Zuo, Chen, Li, Deng, Chen, Behler, Cs\'{a}nyi, Shapeev, Thompson, Wood, and Ong]{Zuo-2020}
Yunxing Zuo, Chi Chen, Xiangguo Li, Zhi Deng, Yiming Chen, J\"{o}rg Behler, G\'{a}bor Cs\'{a}nyi, Alexander~V. Shapeev, Aidan~P. Thompson, Mitchell~A. Wood, and Shyue~Ping Ong.
\newblock Performance and {{Cost Assessment}} of {{Machine Learning Interatomic Potentials}}.
\newblock \emph{The Journal of Physical Chemistry A}, 124\penalty0 (4):\penalty0 731--745, 2020.
\newblock ISSN 1089-5639.
\newblock \doi{10.1021/acs.jpca.9b08723}.
\newblock URL \url{https://doi.org/10.1021/acs.jpca.9b08723}.

\bibitem[Unke et~al.(2021)Unke, Chmiela, Sauceda, Gastegger, Poltavsky, Sch\"{u}tt, Tkatchenko, and M\"{u}ller]{Unke-2021}
Oliver~T. Unke, Stefan Chmiela, Huziel~E. Sauceda, Michael Gastegger, Igor Poltavsky, Kristof~T. Sch\"{u}tt, Alexandre Tkatchenko, and Klaus-Robert M\"{u}ller.
\newblock Machine {{Learning Force Fields}}.
\newblock \emph{Chemical Reviews}, 121\penalty0 (16):\penalty0 10142--10186, 2021.
\newblock ISSN 0009-2665.
\newblock \doi{10.1021/acs.chemrev.0c01111}.
\newblock URL \url{https://doi.org/10.1021/acs.chemrev.0c01111}.

\bibitem[Behler and Parrinello(2007)]{behler2007}
J{\"o}rg Behler and Michele Parrinello.
\newblock Generalized neural-network representation of high-dimensional potential-energy surfaces.
\newblock \emph{Physical Review Letters}, 98\penalty0 (14):\penalty0 146401, 2007.
\newblock \doi{10.1103/PhysRevLett.98.146401}.

\bibitem[Behler(2016)]{Behler-2016}
J\"{o}rg Behler.
\newblock Perspective: {{Machine}} learning potentials for atomistic simulations.
\newblock \emph{The Journal of Chemical Physics}, 145\penalty0 (17):\penalty0 170901, 2016.
\newblock ISSN 0021-9606.
\newblock \doi{10.1063/1.4966192}.
\newblock URL \url{https://doi.org/10.1063/1.4966192}.

\bibitem[Thiemann et~al.(2024-12)Thiemann, O'Neill, Kapil, Michaelides, and Schran]{Thiemann-2024}
Fabian~L Thiemann, Niamh O'Neill, Venkat Kapil, Angelos Michaelides, and Christoph Schran.
\newblock Introduction to machine learning potentials for atomistic simulations.
\newblock \emph{Journal of Physics: Condensed Matter}, 37\penalty0 (7):\penalty0 073002, 2024-12.
\newblock ISSN 0953-8984.
\newblock \doi{10.1088/1361-648X/ad9657}.
\newblock URL \url{https://doi.org/10.1088/1361-648X/ad9657}.

\bibitem[Jacobs et~al.(2025)Jacobs, Morgan, Attarian, Meng, Shen, Wu, Xie, Yang, Artrith, Blaiszik, Ceder, Choudhary, Csanyi, Cubuk, Deng, Drautz, Fu, Godwin, Honavar, Isayev, Johansson, Kozinsky, Martiniani, Ong, Poltavsky, Schmidt, Takamoto, Thompson, Westermayr, and Wood]{Jacobs-2025}
Ryan Jacobs, Dane Morgan, Siamak Attarian, Jun Meng, Chen Shen, Zhenghao Wu, Clare~Yijia Xie, Julia~H. Yang, Nongnuch Artrith, Ben Blaiszik, Gerbrand Ceder, Kamal Choudhary, Gabor Csanyi, Ekin~Dogus Cubuk, Bowen Deng, Ralf Drautz, Xiang Fu, Jonathan Godwin, Vasant Honavar, Olexandr Isayev, Anders Johansson, Boris Kozinsky, Stefano Martiniani, Shyue~Ping Ong, Igor Poltavsky, KJ~Schmidt, So~Takamoto, Aidan~P. Thompson, Julia Westermayr, and Brandon~M. Wood.
\newblock A practical guide to machine learning interatomic potentials – {{Status}} and future.
\newblock \emph{Current Opinion in Solid State and Materials Science}, 35:\penalty0 101214, 2025.
\newblock ISSN 1359-0286.
\newblock \doi{10.1016/j.cossms.2025.101214}.
\newblock URL \url{https://www.sciencedirect.com/science/article/pii/S1359028625000014}.

\bibitem[Friederich et~al.(2021-06)Friederich, H\"{a}se, Proppe, and Aspuru-Guzik]{Friederich-2021}
Pascal Friederich, Florian H\"{a}se, Jonny Proppe, and Al\'{a}n Aspuru-Guzik.
\newblock Machine-learned potentials for next-generation matter simulations.
\newblock \emph{Nature Materials}, 20\penalty0 (6):\penalty0 750--761, 2021-06.
\newblock ISSN 1476-4660.
\newblock \doi{10.1038/s41563-020-0777-6}.
\newblock URL \url{https://www.nature.com/articles/s41563-020-0777-6}.

\bibitem[Deringer et~al.(2019)Deringer, Caro, and Cs\'{a}nyi]{Deringer-2019}
Volker~L. Deringer, Miguel~A. Caro, and G\'{a}bor Cs\'{a}nyi.
\newblock Machine {{Learning Interatomic Potentials}} as {{Emerging Tools}} for {{Materials Science}}.
\newblock \emph{Advanced Materials}, 31\penalty0 (46):\penalty0 1902765, 2019.
\newblock ISSN 1521-4095.
\newblock \doi{10.1002/adma.201902765}.
\newblock URL \url{https://onlinelibrary.wiley.com/doi/abs/10.1002/adma.201902765}.

\bibitem[Chmiela et~al.(2018)Chmiela, Sauceda, M\"{u}ller, and Tkatchenko]{Chmiela-2018}
Stefan Chmiela, Huziel~E. Sauceda, Klaus-Robert M\"{u}ller, and Alexandre Tkatchenko.
\newblock Towards exact molecular dynamics simulations with machine-learned force fields.
\newblock \emph{Nature Communications}, 9\penalty0 (1):\penalty0 3887, 2018.
\newblock ISSN 2041-1723.
\newblock \doi{10.1038/s41467-018-06169-2}.
\newblock URL \url{https://www.nature.com/articles/s41467-018-06169-2}.

\bibitem[von Lilienfeld and Burke(2020)]{vonLilienfeld-2020}
O.~Anatole von Lilienfeld and Kieron Burke.
\newblock Retrospective on a decade of machine learning for chemical discovery.
\newblock \emph{Nature Communications}, 11\penalty0 (1):\penalty0 4895, 2020.
\newblock ISSN 2041-1723.
\newblock \doi{10.1038/s41467-020-18556-9}.
\newblock URL \url{https://www.nature.com/articles/s41467-020-18556-9}.

\bibitem[Smith et~al.(2021)Smith, Nebgen, Mathew, Chen, Lubbers, Burakovsky, Tretiak, Nam, Germann, Fensin, and Barros]{Smith-2021}
Justin~S. Smith, Benjamin Nebgen, Nithin Mathew, Jie Chen, Nicholas Lubbers, Leonid Burakovsky, Sergei Tretiak, Hai~Ah Nam, Timothy Germann, Saryu Fensin, and Kipton Barros.
\newblock Automated discovery of a robust interatomic potential for aluminum.
\newblock \emph{Nature Communications}, 12\penalty0 (1):\penalty0 1257, 2021.
\newblock ISSN 2041-1723.
\newblock \doi{10.1038/s41467-021-21376-0}.
\newblock URL \url{https://www.nature.com/articles/s41467-021-21376-0}.

\bibitem[Musaelian et~al.(2023)Musaelian, Batzner, Johansson, Sun, Owen, Kornbluth, and Kozinsky]{Musaelian-2023}
Albert Musaelian, Simon Batzner, Anders Johansson, Lixin Sun, Cameron~J. Owen, Mordechai Kornbluth, and Boris Kozinsky.
\newblock Learning local equivariant representations for large-scale atomistic dynamics.
\newblock \emph{Nature Communications}, 14\penalty0 (1):\penalty0 579, 2023.
\newblock ISSN 2041-1723.
\newblock \doi{10.1038/s41467-023-36329-y}.
\newblock URL \url{https://www.nature.com/articles/s41467-023-36329-y}.

\bibitem[Lu et~al.(2021)Lu, Wang, Chen, Lin, Car, E, Jia, and Zhang]{Lu-2021}
Denghui Lu, Han Wang, Mohan Chen, Lin Lin, Roberto Car, Weinan E, Weile Jia, and Linfeng Zhang.
\newblock 86 {{PFLOPS Deep Potential Molecular Dynamics}} simulation of 100 million atoms with {\textit{ab initio}} accuracy.
\newblock \emph{Computer Physics Communications}, 259:\penalty0 107624, 2021.
\newblock ISSN 0010-4655.
\newblock \doi{10.1016/j.cpc.2020.107624}.
\newblock URL \url{https://www.sciencedirect.com/science/article/pii/S001046552030299X}.

\bibitem[Fellman et~al.(2025)Fellman, Byggm\"{a}star, Granberg, Nordlund, and Djurabekova]{Fellman-2025}
A.~Fellman, J.~Byggm\"{a}star, F.~Granberg, K.~Nordlund, and F.~Djurabekova.
\newblock Fast and accurate machine-learned interatomic potentials for large-scale simulations of {{Cu}}, {{Al}}, and {{Ni}}.
\newblock \emph{Physical Review Materials}, 9\penalty0 (5):\penalty0 053807, 2025.
\newblock \doi{10.1103/PhysRevMaterials.9.053807}.
\newblock URL \url{https://link.aps.org/doi/10.1103/PhysRevMaterials.9.053807}.

\bibitem[Nguyen-Cong et~al.(2021)Nguyen-Cong, Willman, Moore, Belonoshko, Gayatri, Weinberg, Wood, Thompson, and Oleynik]{Nguyen-Cong-2021}
Kien Nguyen-Cong, Jonathan~T. Willman, Stan~G. Moore, Anatoly~B. Belonoshko, Rahulkumar Gayatri, Evan Weinberg, Mitchell~A. Wood, Aidan~P. Thompson, and Ivan~I. Oleynik.
\newblock Billion atom molecular dynamics simulations of carbon at extreme conditions and experimental time and length scales.
\newblock In \emph{Proceedings of the {{International Conference}} for {{High Performance Computing}}, {{Networking}}, {{Storage}} and {{Analysis}}}, {{SC}} '21, pages 1--12, New York, NY, USA, 2021. Association for Computing Machinery.
\newblock ISBN 978-1-4503-8442-1.
\newblock \doi{10.1145/3458817.3487400}.
\newblock URL \url{https://dl.acm.org/doi/10.1145/3458817.3487400}.

\bibitem[Jia et~al.(2020)Jia, Wang, Chen, Lu, Lin, Car, E, and Zhang]{Jia-2020}
Weile Jia, Han Wang, Mohan Chen, Denghui Lu, Lin Lin, Roberto Car, Weinan E, and Linfeng Zhang.
\newblock Pushing the limit of molecular dynamics with ab initio accuracy to 100 million atoms with machine learning.
\newblock In \emph{Proceedings of the {{International Conference}} for {{High Performance Computing}}, {{Networking}}, {{Storage}} and {{Analysis}}}, {{SC}} '20, pages 1--14, Atlanta, Georgia, 2020. IEEE Press.
\newblock ISBN 978-1-7281-9998-6.

\bibitem[Deringer et~al.(2021-01)Deringer, Bernstein, Cs\'{a}nyi, Ben~Mahmoud, Ceriotti, Wilson, Drabold, and Elliott]{Deringer-2021}
Volker~L. Deringer, Noam Bernstein, G\'{a}bor Cs\'{a}nyi, Chiheb Ben~Mahmoud, Michele Ceriotti, Mark Wilson, David~A. Drabold, and Stephen~R. Elliott.
\newblock Origins of structural and electronic transitions in disordered silicon.
\newblock \emph{Nature}, 589\penalty0 (7840):\penalty0 59--64, 2021-01.
\newblock ISSN 1476-4687.
\newblock \doi{10.1038/s41586-020-03072-z}.
\newblock URL \url{https://www.nature.com/articles/s41586-020-03072-z}.

\bibitem[Wang et~al.(2024-11)Wang, He, Li, Li, Bi, Wang, Cheng, Shen, Meng, Zhang, Liu, Wang, Li, Shao, and Liu]{Wang-2024}
Tong Wang, Xinheng He, Mingyu Li, Yatao Li, Ran Bi, Yusong Wang, Chaoran Cheng, Xiangzhen Shen, Jiawei Meng, He~Zhang, Haiguang Liu, Zun Wang, Shaoning Li, Bin Shao, and Tie-Yan Liu.
\newblock Ab initio characterization of protein molecular dynamics with {{AI2BMD}}.
\newblock \emph{Nature}, 635\penalty0 (8040):\penalty0 1019--1027, 2024-11.
\newblock ISSN 1476-4687.
\newblock \doi{10.1038/s41586-024-08127-z}.
\newblock URL \url{https://www.nature.com/articles/s41586-024-08127-z}.

\bibitem[Unke et~al.(2024)Unke, St\"{o}hr, Ganscha, Unterthiner, Maennel, Kashubin, Ahlin, Gastegger, Medrano~Sandonas, Berryman, Tkatchenko, and M\"{u}ller]{Unke-2024}
Oliver~T. Unke, Martin St\"{o}hr, Stefan Ganscha, Thomas Unterthiner, Hartmut Maennel, Sergii Kashubin, Daniel Ahlin, Michael Gastegger, Leonardo Medrano~Sandonas, Joshua~T. Berryman, Alexandre Tkatchenko, and Klaus-Robert M\"{u}ller.
\newblock Biomolecular dynamics with machine-learned quantum-mechanical force fields trained on diverse chemical fragments.
\newblock \emph{Science Advances}, 10\penalty0 (14):\penalty0 eadn4397, 2024.
\newblock \doi{10.1126/sciadv.adn4397}.
\newblock URL \url{https://www.science.org/doi/10.1126/sciadv.adn4397}.

\bibitem[Kozinsky et~al.(2023)Kozinsky, Musaelian, Johansson, and Batzner]{Kozinsky-2023}
Boris Kozinsky, Albert Musaelian, Anders Johansson, and Simon Batzner.
\newblock Scaling the {{Leading Accuracy}} of {{Deep Equivariant Models}} to {{Biomolecular Simulations}} of {{Realistic Size}}.
\newblock In \emph{Proceedings of the {{International Conference}} for {{High Performance Computing}}, {{Networking}}, {{Storage}} and {{Analysis}}}, {{SC}} '23, pages 1--12, New York, NY, USA, 2023. Association for Computing Machinery.
\newblock ISBN 979-8-4007-0109-2.
\newblock \doi{10.1145/3581784.3627041}.
\newblock URL \url{https://dl.acm.org/doi/10.1145/3581784.3627041}.

\bibitem[Chen et~al.(2023{\natexlab{a}})Chen, Shang, and Liu]{Chen-2023}
Dongxiao Chen, Cheng Shang, and Zhi-Pan Liu.
\newblock Machine-learning atomic simulation for heterogeneous catalysis.
\newblock \emph{npj Computational Materials}, 9\penalty0 (1):\penalty0 2, 2023{\natexlab{a}}.
\newblock ISSN 2057-3960.
\newblock \doi{10.1038/s41524-022-00959-5}.
\newblock URL \url{https://www.nature.com/articles/s41524-022-00959-5}.

\bibitem[Zhang et~al.(2024{\natexlab{a}})Zhang, Juraskova, and Duarte]{Zhang-2024}
Hanwen Zhang, Veronika Juraskova, and Fernanda Duarte.
\newblock Modelling chemical processes in explicit solvents with machine learning potentials.
\newblock \emph{Nature Communications}, 15\penalty0 (1):\penalty0 6114, 2024{\natexlab{a}}.
\newblock ISSN 2041-1723.
\newblock \doi{10.1038/s41467-024-50418-6}.
\newblock URL \url{https://www.nature.com/articles/s41467-024-50418-6}.

\bibitem[Wander et~al.(2025)Wander, Shuaibi, Kitchin, Ulissi, and Zitnick]{Wander-2025}
Brook Wander, Muhammed Shuaibi, John~R. Kitchin, Zachary~W. Ulissi, and C.~Lawrence Zitnick.
\newblock {{CatTSunami}}: {{Accelerating Transition State Energy Calculations}} with {{Pretrained Graph Neural Networks}}.
\newblock \emph{ACS Catalysis}, 15\penalty0 (7):\penalty0 5283--5294, 2025.
\newblock \doi{10.1021/acscatal.4c04272}.
\newblock URL \url{https://doi.org/10.1021/acscatal.4c04272}.

\bibitem[Zhang et~al.(2024-05)Zhang, Mako\'{s}, Jadrich, Kraka, Barros, Nebgen, Tretiak, Isayev, Lubbers, Messerly, and Smith]{Zhang-2024a}
Shuhao Zhang, Ma\l{}gorzata~Z. Mako\'{s}, Ryan~B. Jadrich, Elfi Kraka, Kipton Barros, Benjamin~T. Nebgen, Sergei Tretiak, Olexandr Isayev, Nicholas Lubbers, Richard~A. Messerly, and Justin~S. Smith.
\newblock Exploring the frontiers of condensed-phase chemistry with a general reactive machine learning potential.
\newblock \emph{Nature Chemistry}, 16\penalty0 (5):\penalty0 727--734, 2024-05.
\newblock ISSN 1755-4349.
\newblock \doi{10.1038/s41557-023-01427-3}.
\newblock URL \url{https://www.nature.com/articles/s41557-023-01427-3}.

\bibitem[Guan et~al.(2023-11)Guan, Heindel, Ko, Yang, and Head-Gordon]{Guan-2023}
Xingyi Guan, Joseph~P. Heindel, Taehee Ko, Chao Yang, and Teresa Head-Gordon.
\newblock Using machine learning to go beyond potential energy surface benchmarking for chemical reactivity.
\newblock \emph{Nature Computational Science}, 3\penalty0 (11):\penalty0 965--974, 2023-11.
\newblock ISSN 2662-8457.
\newblock \doi{10.1038/s43588-023-00549-5}.
\newblock URL \url{https://www.nature.com/articles/s43588-023-00549-5}.

\bibitem[Chen et~al.(2023{\natexlab{b}})Chen, Zhang, and Zhang]{Chen-2023a}
Benjamin W.~J. Chen, Xinglong Zhang, and Jia Zhang.
\newblock Accelerating explicit solvent models of heterogeneous catalysts with machine learning interatomic potentials.
\newblock \emph{Chemical Science}, 14\penalty0 (31):\penalty0 8338--8354, 2023{\natexlab{b}}.
\newblock ISSN 2041-6539.
\newblock \doi{10.1039/D3SC02482B}.
\newblock URL \url{https://pubs.rsc.org/en/content/articlelanding/2023/sc/d3sc02482b}.

\bibitem[Pandey et~al.(2022-03)Pandey, Fernandez, Gentile, Isayev, Tropsha, Stern, and Cherkasov]{Pandey-2022}
Mohit Pandey, Michael Fernandez, Francesco Gentile, Olexandr Isayev, Alexander Tropsha, Abraham~C. Stern, and Artem Cherkasov.
\newblock The transformational role of {{GPU}} computing and deep learning in drug discovery.
\newblock \emph{Nature Machine Intelligence}, 4\penalty0 (3):\penalty0 211--221, 2022-03.
\newblock ISSN 2522-5839.
\newblock \doi{10.1038/s42256-022-00463-x}.
\newblock URL \url{https://www.nature.com/articles/s42256-022-00463-x}.

\bibitem[Stone et~al.(2007)Stone, Phillips, Freddolino, Hardy, Trabuco, and Schulten]{Stone-2007}
John~E. Stone, James~C. Phillips, Lydia Freddolino, David~J. Hardy, Leonardo~G. Trabuco, and Klaus Schulten.
\newblock Accelerating molecular modeling applications with graphics processors.
\newblock \emph{Journal of Computational Chemistry}, 28\penalty0 (16):\penalty0 2618--2640, 2007.
\newblock ISSN 1096-987X.
\newblock \doi{10.1002/jcc.20829}.
\newblock URL \url{https://onlinelibrary.wiley.com/doi/abs/10.1002/jcc.20829}.

\bibitem[Stone et~al.(2010)Stone, Hardy, Ufimtsev, and Schulten]{Stone-2010}
John~E. Stone, David~J. Hardy, Ivan~S. Ufimtsev, and Klaus Schulten.
\newblock {{GPU-accelerated}} molecular modeling coming of age.
\newblock \emph{Journal of Molecular Graphics and Modelling}, 29\penalty0 (2):\penalty0 116--125, 2010.
\newblock ISSN 1093-3263.
\newblock \doi{10.1016/j.jmgm.2010.06.010}.
\newblock URL \url{https://www.sciencedirect.com/science/article/pii/S1093326310000914}.

\bibitem[Sch\"{u}tt et~al.(2017)Sch\"{u}tt, Kindermans, Sauceda, Chmiela, Tkatchenko, and M\"{u}ller]{Schütt-2017}
Kristof~T. Sch\"{u}tt, Pieter-Jan Kindermans, Huziel~E. Sauceda, Stefan Chmiela, Alexandre Tkatchenko, and Klaus-Robert M\"{u}ller.
\newblock {{SchNet}}: {{A}} continuous-filter convolutional neural network for modeling quantum interactions, 2017.
\newblock URL \url{http://arxiv.org/abs/1706.08566}.

\bibitem[Gilmer et~al.(2017)Gilmer, Schoenholz, Riley, Vinyals, and Dahl]{Gilmer-2017}
Justin Gilmer, Samuel~S. Schoenholz, Patrick~F. Riley, Oriol Vinyals, and George~E. Dahl.
\newblock Neural message passing for {{Quantum}} chemistry.
\newblock In \emph{Proceedings of the 34th {{International Conference}} on {{Machine Learning}} - {{Volume}} 70}, {{ICML}}'17, pages 1263--1272, Sydney, NSW, Australia, 2017. JMLR.org.
\newblock URL \url{https://dl.acm.org/doi/10.5555/3305381.3305512}.

\bibitem[Batzner et~al.(2022)Batzner, Musaelian, Sun, Geiger, Mailoa, Kornbluth, Molinari, Smidt, and Kozinsky]{Batzner-2022}
Simon Batzner, Albert Musaelian, Lixin Sun, Mario Geiger, Jonathan~P. Mailoa, Mordechai Kornbluth, Nicola Molinari, Tess~E. Smidt, and Boris Kozinsky.
\newblock E(3)-equivariant graph neural networks for data-efficient and accurate interatomic potentials.
\newblock \emph{Nature Communications}, 13\penalty0 (1):\penalty0 2453, 2022.
\newblock ISSN 2041-1723.
\newblock \doi{10.1038/s41467-022-29939-5}.
\newblock URL \url{https://www.nature.com/articles/s41467-022-29939-5}.

\bibitem[Batatia et~al.(2022)Batatia, Kovacs, Simm, Ortner, and Csanyi]{Batatia-2022}
Ilyes Batatia, David~P. Kovacs, Gregor Simm, Christoph Ortner, and Gabor Csanyi.
\newblock {{MACE}}: {{Higher Order Equivariant Message Passing Neural Networks}} for {{Fast}} and {{Accurate Force Fields}}.
\newblock \emph{Advances in Neural Information Processing Systems}, 35:\penalty0 11423--11436, 2022.
\newblock URL \url{https://proceedings.neurips.cc/paper/2022/file/4a36c3c5b668f4de3d5de4b37c94a823-Paper-Conference.pdf}.

\bibitem[Mazitov et~al.(2025)Mazitov, Bigi, Kellner, Pegolo, Tisi, Fraux, Pozdnyakov, Loche, and Ceriotti]{Mazitov-2025}
Arslan Mazitov, Filippo Bigi, Matthias Kellner, Paolo Pegolo, Davide Tisi, Guillaume Fraux, Sergey Pozdnyakov, Philip Loche, and Michele Ceriotti.
\newblock {{PET-MAD}} as a lightweight universal interatomic potential for advanced materials modeling.
\newblock \emph{Nature Communications}, 16\penalty0 (1):\penalty0 10653, 2025.
\newblock ISSN 2041-1723.
\newblock \doi{10.1038/s41467-025-65662-7}.
\newblock URL \url{https://www.nature.com/articles/s41467-025-65662-7}.

\bibitem[Barroso-Luque et~al.(2024)Barroso-Luque, Shuaibi, Fu, Wood, Dzamba, Gao, Rizvi, Zitnick, and Ulissi]{Barroso-Luque-2024}
Luis Barroso-Luque, Muhammed Shuaibi, Xiang Fu, Brandon~M. Wood, Misko Dzamba, Meng Gao, Ammar Rizvi, C.~Lawrence Zitnick, and Zachary~W. Ulissi.
\newblock Open {{Materials}} 2024 ({{OMat24}}) {{Inorganic Materials Dataset}} and {{Models}}, 2024.
\newblock URL \url{http://arxiv.org/abs/2410.12771}.

\bibitem[Chanussot et~al.(2021)Chanussot, Das, Goyal, Lavril, Shuaibi, Riviere, Tran, Heras-Domingo, Ho, Hu, Palizhati, Sriram, Wood, Yoon, Parikh, Zitnick, and Ulissi]{Chanussot-2021}
Lowik Chanussot, Abhishek Das, Siddharth Goyal, Thibaut Lavril, Muhammed Shuaibi, Morgane Riviere, Kevin Tran, Javier Heras-Domingo, Caleb Ho, Weihua Hu, Aini Palizhati, Anuroop Sriram, Brandon Wood, Junwoong Yoon, Devi Parikh, C.~Lawrence Zitnick, and Zachary Ulissi.
\newblock Open {{Catalyst}} 2020 ({{OC20}}) {{Dataset}} and {{Community Challenges}}.
\newblock \emph{ACS Catalysis}, 11\penalty0 (10):\penalty0 6059--6072, 2021.
\newblock \doi{10.1021/acscatal.0c04525}.
\newblock URL \url{https://doi.org/10.1021/acscatal.0c04525}.

\bibitem[Gharakhanyan et~al.(2025)Gharakhanyan, Barroso-Luque, Yang, Shuaibi, Michel, Levine, Dzamba, Fu, Gao, Liu, Ni, Noori, Wood, Uyttendaele, Boromand, Zitnick, Marom, Ulissi, and Sriram]{Gharakhanyan-2025}
Vahe Gharakhanyan, Luis Barroso-Luque, Yi~Yang, Muhammed Shuaibi, Kyle Michel, Daniel~S. Levine, Misko Dzamba, Xiang Fu, Meng Gao, Xingyu Liu, Haoran Ni, Keian Noori, Brandon~M. Wood, Matt Uyttendaele, Arman Boromand, C.~Lawrence Zitnick, Noa Marom, Zachary~W. Ulissi, and Anuroop Sriram.
\newblock Open {{Molecular Crystals}} 2025 ({{OMC25}}) {{Dataset}} and {{Models}}, 2025.
\newblock URL \url{http://arxiv.org/abs/2508.02651}.

\bibitem[Levine et~al.(2025)Levine, Shuaibi, Spotte-Smith, Taylor, Hasyim, Michel, Batatia, Cs\'{a}nyi, Dzamba, Eastman, Frey, Fu, Gharakhanyan, Krishnapriyan, Rackers, Raja, Rizvi, Rosen, Ulissi, Vargas, Zitnick, Blau, and Wood]{Levine-2025}
Daniel~S. Levine, Muhammed Shuaibi, Evan Walter~Clark Spotte-Smith, Michael~G. Taylor, Muhammad~R. Hasyim, Kyle Michel, Ilyes Batatia, G\'{a}bor Cs\'{a}nyi, Misko Dzamba, Peter Eastman, Nathan~C. Frey, Xiang Fu, Vahe Gharakhanyan, Aditi~S. Krishnapriyan, Joshua~A. Rackers, Sanjeev Raja, Ammar Rizvi, Andrew~S. Rosen, Zachary Ulissi, Santiago Vargas, C.~Lawrence Zitnick, Samuel~M. Blau, and Brandon~M. Wood.
\newblock The {{Open Molecules}} 2025 ({{OMol25}}) {{Dataset}}, {{Evaluations}}, and {{Models}}, 2025.
\newblock URL \url{http://arxiv.org/abs/2505.08762}.

\bibitem[Sahoo et~al.(2025)Sahoo, Maraschin, Levine, Ulissi, Zitnick, Varley, Gauthier, Govindarajan, and Shuaibi]{Sahoo-2025}
Sushree~Jagriti Sahoo, Mikael Maraschin, Daniel~S. Levine, Zachary Ulissi, C.~Lawrence Zitnick, Joel~B. Varley, Joseph~A. Gauthier, Nitish Govindarajan, and Muhammed Shuaibi.
\newblock The {{Open Catalyst}} 2025 ({{OC25}}) {{Dataset}} and {{Models}} for {{Solid-Liquid Interfaces}}, 2025.
\newblock URL \url{http://arxiv.org/abs/2509.17862}.

\bibitem[Sriram et~al.(2024)Sriram, Choi, Yu, Brabson, Das, Ulissi, Uyttendaele, Medford, and Sholl]{Sriram-2024}
Anuroop Sriram, Sihoon Choi, Xiaohan Yu, Logan~M. Brabson, Abhishek Das, Zachary Ulissi, Matt Uyttendaele, Andrew~J. Medford, and David~S. Sholl.
\newblock The {{Open DAC}} 2023 {{Dataset}} and {{Challenges}} for {{Sorbent Discovery}} in {{Direct Air Capture}}.
\newblock \emph{ACS Central Science}, 10\penalty0 (5):\penalty0 923--941, 2024.
\newblock ISSN 2374-7943.
\newblock \doi{10.1021/acscentsci.3c01629}.
\newblock URL \url{https://doi.org/10.1021/acscentsci.3c01629}.

\bibitem[Tran et~al.(2023)Tran, Lan, Shuaibi, Wood, Goyal, Das, Heras-Domingo, Kolluru, Rizvi, Shoghi, Sriram, Therrien, Abed, Voznyy, Sargent, Ulissi, and Zitnick]{Tran-2023}
Richard Tran, Janice Lan, Muhammed Shuaibi, Brandon~M. Wood, Siddharth Goyal, Abhishek Das, Javier Heras-Domingo, Adeesh Kolluru, Ammar Rizvi, Nima Shoghi, Anuroop Sriram, F\'{e}lix Therrien, Jehad Abed, Oleksandr Voznyy, Edward~H. Sargent, Zachary Ulissi, and C.~Lawrence Zitnick.
\newblock The {{Open Catalyst}} 2022 ({{OC22}}) {{Dataset}} and {{Challenges}} for {{Oxide Electrocatalysts}}.
\newblock \emph{ACS Catalysis}, 13\penalty0 (5):\penalty0 3066--3084, 2023.
\newblock \doi{10.1021/acscatal.2c05426}.
\newblock URL \url{https://doi.org/10.1021/acscatal.2c05426}.

\bibitem[Deng et~al.(2023-09)Deng, Zhong, Jun, Riebesell, Han, Bartel, and Ceder]{Deng-2023}
Bowen Deng, Peichen Zhong, KyuJung Jun, Janosh Riebesell, Kevin Han, Christopher~J. Bartel, and Gerbrand Ceder.
\newblock {{CHGNet}} as a pretrained universal neural network potential for charge-informed atomistic modelling.
\newblock \emph{Nature Machine Intelligence}, 5\penalty0 (9):\penalty0 1031--1041, 2023-09.
\newblock ISSN 2522-5839.
\newblock \doi{10.1038/s42256-023-00716-3}.
\newblock URL \url{https://www.nature.com/articles/s42256-023-00716-3}.

\bibitem[Devereux et~al.(2020)Devereux, Smith, Huddleston, Barros, Zubatyuk, Isayev, and Roitberg]{Devereux-2020}
Christian Devereux, Justin~S. Smith, Kate~K. Huddleston, Kipton Barros, Roman Zubatyuk, Olexandr Isayev, and Adrian~E. Roitberg.
\newblock Extending the {{Applicability}} of the {{ANI Deep Learning Molecular Potential}} to {{Sulfur}} and {{Halogens}}.
\newblock \emph{Journal of Chemical Theory and Computation}, 16\penalty0 (7):\penalty0 4192--4202, 2020.
\newblock ISSN 1549-9618, 1549-9626.
\newblock \doi{10.1021/acs.jctc.0c00121}.
\newblock URL \url{https://pubs.acs.org/doi/10.1021/acs.jctc.0c00121}.

\bibitem[Eastman et~al.(2023)Eastman, Behara, Dotson, Galvelis, Herr, Horton, Mao, Chodera, Pritchard, Wang, De~Fabritiis, and Markland]{Eastman-2023}
Peter Eastman, Pavan~Kumar Behara, David~L. Dotson, Raimondas Galvelis, John~E. Herr, Josh~T. Horton, Yuezhi Mao, John~D. Chodera, Benjamin~P. Pritchard, Yuanqing Wang, Gianni De~Fabritiis, and Thomas~E. Markland.
\newblock {{SPICE}}, {{A Dataset}} of {{Drug-like Molecules}} and {{Peptides}} for {{Training Machine Learning Potentials}}.
\newblock \emph{Scientific Data}, 10\penalty0 (1):\penalty0 11, 2023.
\newblock ISSN 2052-4463.
\newblock \doi{10.1038/s41597-022-01882-6}.
\newblock URL \url{https://www.nature.com/articles/s41597-022-01882-6}.

\bibitem[Schmidt et~al.(2024)Schmidt, Cerqueira, Romero, Loew, J\"{a}ger, Wang, Botti, and Marques]{Schmidt-2024}
Jonathan Schmidt, Tiago F.~T. Cerqueira, Aldo~H. Romero, Antoine Loew, Fabian J\"{a}ger, Hai-Chen Wang, Silvana Botti, and Miguel A.~L. Marques.
\newblock Improving machine-learning models in materials science through large datasets.
\newblock \emph{Materials Today Physics}, 48:\penalty0 101560, 2024.
\newblock ISSN 2542-5293.
\newblock \doi{10.1016/j.mtphys.2024.101560}.
\newblock URL \url{https://www.sciencedirect.com/science/article/pii/S2542529324002360}.

\bibitem[Huang et~al.(2025)Huang, Deng, Zhong, Kaplan, Persson, and Ceder]{Huang-2025}
Xu~Huang, Bowen Deng, Peichen Zhong, Aaron~D. Kaplan, Kristin~A. Persson, and Gerbrand Ceder.
\newblock Cross-functional transferability in foundation machine learning interatomic potentials.
\newblock \emph{npj Computational Materials}, 11\penalty0 (1):\penalty0 313, 2025.
\newblock ISSN 2057-3960.
\newblock \doi{10.1038/s41524-025-01796-y}.
\newblock URL \url{https://www.nature.com/articles/s41524-025-01796-y}.

\bibitem[Batatia et~al.(2025)Batatia, Benner, Chiang, Elena, Kov\'{a}cs, Riebesell, Advincula, Asta, Avaylon, Baldwin, Berger, Bernstein, Bhowmik, Bigi, Blau, C\u{a}rare, Ceriotti, Chong, Darby, De, Della~Pia, Deringer, Elijo\v{s}ius, El-Machachi, Fako, Falcioni, Ferrari, Gardner, Gawkowski, Genreith-Schriever, George, Goodall, Grandel, Grey, Grigorev, Han, Handley, Heenen, Hermansson, Ho, Hofmann, Holm, Jaafar, Jakob, Jung, Kapil, Kaplan, Karimitari, Kermode, Kourtis, Kroupa, Kullgren, Kuner, Kuryla, Liepuoniute, Lin, Margraf, Magd\u{a}u, Michaelides, Moore, Naik, Niblett, Norwood, O'Neill, Ortner, Persson, Reuter, Rosen, Rosset, Schaaf, Schran, Shi, Sivonxay, Stenczel, Sutton, Svahn, Swinburne, Tilly, van~der Oord, Vargas, Varga-Umbrich, Vegge, Vondr\'{a}k, Wang, Witt, Wolf, Zills, and Cs\'{a}nyi]{Batatia-2025}
Ilyes Batatia, Philipp Benner, Yuan Chiang, Alin~M. Elena, D\'{a}vid~P. Kov\'{a}cs, Janosh Riebesell, Xavier~R. Advincula, Mark Asta, Matthew Avaylon, William~J. Baldwin, Fabian Berger, Noam Bernstein, Arghya Bhowmik, Filippo Bigi, Samuel~M. Blau, Vlad C\u{a}rare, Michele Ceriotti, Sanggyu Chong, James~P. Darby, Sandip De, Flaviano Della~Pia, Volker~L. Deringer, Rokas Elijo\v{s}ius, Zakariya El-Machachi, Edvin Fako, Fabio Falcioni, Andrea~C. Ferrari, John L.~A. Gardner, Miko\l{}aj~J. Gawkowski, Annalena Genreith-Schriever, Janine George, Rhys E.~A. Goodall, Jonas Grandel, Clare~P. Grey, Petr Grigorev, Shuang Han, Will Handley, Hendrik~H. Heenen, Kersti Hermansson, Cheuk~Hin Ho, Stephan Hofmann, Christian Holm, Jad Jaafar, Konstantin~S. Jakob, Hyunwook Jung, Venkat Kapil, Aaron~D. Kaplan, Nima Karimitari, James~R. Kermode, Panagiotis Kourtis, Namu Kroupa, Jolla Kullgren, Matthew~C. Kuner, Domantas Kuryla, Guoda Liepuoniute, Chen Lin, Johannes~T. Margraf, Ioan-Bogdan Magd\u{a}u, Angelos Michaelides, J.~Harry
  Moore, Aakash~A. Naik, Samuel~P. Niblett, Sam~Walton Norwood, Niamh O'Neill, Christoph Ortner, Kristin~A. Persson, Karsten Reuter, Andrew~S. Rosen, Louise A.~M. Rosset, Lars~L. Schaaf, Christoph Schran, Benjamin~X. Shi, Eric Sivonxay, Tam\'{a}s~K. Stenczel, Christopher Sutton, Viktor Svahn, Thomas~D. Swinburne, Jules Tilly, Cas van~der Oord, Santiago Vargas, Eszter Varga-Umbrich, Tejs Vegge, Martin Vondr\'{a}k, Yangshuai Wang, William~C. Witt, Thomas Wolf, Fabian Zills, and G\'{a}bor Cs\'{a}nyi.
\newblock A foundation model for atomistic materials chemistry.
\newblock \emph{The Journal of Chemical Physics}, 163\penalty0 (18):\penalty0 184110, 2025.
\newblock ISSN 0021-9606.
\newblock \doi{10.1063/5.0297006}.
\newblock URL \url{https://doi.org/10.1063/5.0297006}.

\bibitem[Wood et~al.(2025)Wood, Dzamba, Fu, Gao, Shuaibi, Barroso-Luque, Abdelmaqsoud, Gharakhanyan, Kitchin, Levine, Michel, Sriram, Cohen, Das, Rizvi, Sahoo, Ulissi, and Zitnick]{Wood-2025}
Brandon~M. Wood, Misko Dzamba, Xiang Fu, Meng Gao, Muhammed Shuaibi, Luis Barroso-Luque, Kareem Abdelmaqsoud, Vahe Gharakhanyan, John~R. Kitchin, Daniel~S. Levine, Kyle Michel, Anuroop Sriram, Taco Cohen, Abhishek Das, Ammar Rizvi, Sushree~Jagriti Sahoo, Zachary~W. Ulissi, and C.~Lawrence Zitnick.
\newblock {{UMA}}: {{A Family}} of {{Universal Models}} for {{Atoms}}, 2025.
\newblock URL \url{http://arxiv.org/abs/2506.23971}.

\bibitem[Chen and Ong(2022-11)]{Chen-2022}
Chi Chen and Shyue~Ping Ong.
\newblock A universal graph deep learning interatomic potential for the periodic table.
\newblock \emph{Nature Computational Science}, 2\penalty0 (11):\penalty0 718--728, 2022-11.
\newblock ISSN 2662-8457.
\newblock \doi{10.1038/s43588-022-00349-3}.
\newblock URL \url{https://www.nature.com/articles/s43588-022-00349-3}.

\bibitem[Merchant et~al.(2023-12)Merchant, Batzner, Schoenholz, Aykol, Cheon, and Cubuk]{Merchant-2023}
Amil Merchant, Simon Batzner, Samuel~S. Schoenholz, Muratahan Aykol, Gowoon Cheon, and Ekin~Dogus Cubuk.
\newblock Scaling deep learning for materials discovery.
\newblock \emph{Nature}, 624\penalty0 (7990):\penalty0 80--85, 2023-12.
\newblock ISSN 1476-4687.
\newblock \doi{10.1038/s41586-023-06735-9}.
\newblock URL \url{https://www.nature.com/articles/s41586-023-06735-9}.

\bibitem[Zhang et~al.(2024{\natexlab{b}})Zhang, Liu, Zhang, Zhang, Cai, Bi, Du, Qin, Peng, Huang, Li, Shan, Zeng, Zhang, Liu, Li, Chang, Wang, Zhou, Liu, Luo, Wang, Jiang, Wu, Yang, Yang, Yang, Gong, Zhang, Shi, Dai, York, Liu, Zhu, Zhong, Lv, Cheng, Jia, Chen, Ke, E, Zhang, and Wang]{Zhang-2024b}
Duo Zhang, Xinzijian Liu, Xiangyu Zhang, Chengqian Zhang, Chun Cai, Hangrui Bi, Yiming Du, Xuejian Qin, Anyang Peng, Jiameng Huang, Bowen Li, Yifan Shan, Jinzhe Zeng, Yuzhi Zhang, Siyuan Liu, Yifan Li, Junhan Chang, Xinyan Wang, Shuo Zhou, Jianchuan Liu, Xiaoshan Luo, Zhenyu Wang, Wanrun Jiang, Jing Wu, Yudi Yang, Jiyuan Yang, Manyi Yang, Fu-Qiang Gong, Linshuang Zhang, Mengchao Shi, Fu-Zhi Dai, Darrin~M. York, Shi Liu, Tong Zhu, Zhicheng Zhong, Jian Lv, Jun Cheng, Weile Jia, Mohan Chen, Guolin Ke, Weinan E, Linfeng Zhang, and Han Wang.
\newblock {{DPA-2}}: A large atomic model as a multi-task learner.
\newblock \emph{npj Computational Materials}, 10\penalty0 (1):\penalty0 293, 2024{\natexlab{b}}.
\newblock ISSN 2057-3960.
\newblock \doi{10.1038/s41524-024-01493-2}.
\newblock URL \url{https://www.nature.com/articles/s41524-024-01493-2}.

\bibitem[Yang et~al.(2024)Yang, Hu, Zhou, Liu, Shi, Li, Li, Chen, Chen, Zeni, Horton, Pinsler, Fowler, Z\"{u}gner, Xie, Smith, Sun, Wang, Kong, Liu, Hao, and Lu]{Yang-2024}
Han Yang, Chenxi Hu, Yichi Zhou, Xixian Liu, Yu~Shi, Jielan Li, Guanzhi Li, Zekun Chen, Shuizhou Chen, Claudio Zeni, Matthew Horton, Robert Pinsler, Andrew Fowler, Daniel Z\"{u}gner, Tian Xie, Jake Smith, Lixin Sun, Qian Wang, Lingyu Kong, Chang Liu, Hongxia Hao, and Ziheng Lu.
\newblock {{MatterSim}}: {{A Deep Learning Atomistic Model Across Elements}}, {{Temperatures}} and {{Pressures}}, 2024.
\newblock URL \url{http://arxiv.org/abs/2405.04967}.

\bibitem[Radova et~al.(2025)Radova, Stark, Allen, Maurer, and Bart\'{o}k]{Radova-2025}
Mariia Radova, Wojciech~G. Stark, Connor~S. Allen, Reinhard~J. Maurer, and Albert~P. Bart\'{o}k.
\newblock Fine-tuning foundation models of materials interatomic potentials with frozen transfer learning, 2025.
\newblock URL \url{http://arxiv.org/abs/2502.15582}.

\bibitem[Focassio et~al.(2025)Focassio, M.~Freitas, and Schleder]{Focassio-2025}
Bruno Focassio, Luis~Paulo M.~Freitas, and Gabriel~R. Schleder.
\newblock Performance {{Assessment}} of {{Universal Machine Learning Interatomic Potentials}}: {{Challenges}} and {{Directions}} for {{Materials}}' {{Surfaces}}.
\newblock \emph{ACS Applied Materials \& Interfaces}, 17\penalty0 (9):\penalty0 13111--13121, 2025.
\newblock ISSN 1944-8244.
\newblock \doi{10.1021/acsami.4c03815}.
\newblock URL \url{https://doi.org/10.1021/acsami.4c03815}.

\bibitem[Liu et~al.(2025)Liu, Zeng, Luo, Wang, Zhao, and Xu]{Liu-2025}
Xiaoqing Liu, Kehan Zeng, Zedong Luo, Yangshuai Wang, Teng Zhao, and Zhenli Xu.
\newblock Fine-{{Tuning Universal Machine-Learned Interatomic Potentials}}: {{A Tutorial}} on {{Methods}} and {{Applications}}, 2025.
\newblock URL \url{http://arxiv.org/abs/2506.21935}.

\bibitem[Pia et~al.(2025)Pia, X.~Shi, Kapil, Zen, Alf\`{e}, and Michaelides]{Pia-2025}
Flaviano~Della Pia, Benjamin X.~Shi, Venkat Kapil, Andrea Zen, Dario Alf\`{e}, and Angelos Michaelides.
\newblock Accurate and efficient machine learning interatomic potentials for finite temperature modelling of molecular crystals.
\newblock \emph{Chemical Science}, 16\penalty0 (25):\penalty0 11419--11433, 2025.
\newblock \doi{10.1039/D5SC01325A}.
\newblock URL \url{https://pubs.rsc.org/en/content/articlelanding/2025/sc/d5sc01325a}.

\bibitem[Riebesell et~al.(2025-06)Riebesell, Goodall, Benner, Chiang, Deng, Ceder, Asta, Lee, Jain, and Persson]{Riebesell-2025}
Janosh Riebesell, Rhys E.~A. Goodall, Philipp Benner, Yuan Chiang, Bowen Deng, Gerbrand Ceder, Mark Asta, Alpha~A. Lee, Anubhav Jain, and Kristin~A. Persson.
\newblock A framework to evaluate machine learning crystal stability predictions.
\newblock \emph{Nature Machine Intelligence}, 7\penalty0 (6):\penalty0 836--847, 2025-06.
\newblock ISSN 2522-5839.
\newblock \doi{10.1038/s42256-025-01055-1}.
\newblock URL \url{https://www.nature.com/articles/s42256-025-01055-1}.

\bibitem[Chiang et~al.(2025)Chiang, Kreiman, Zhang, Kuner, Weaver, Amin, Park, Lim, Kim, Chrzan, Walsh, Blau, Asta, and Krishnapriyan]{Chiang-2025}
Yuan Chiang, Tobias Kreiman, Christine Zhang, Matthew~C. Kuner, Elizabeth~Jin Weaver, Ishan Amin, Hyunsoo Park, Yunsung Lim, Jihan Kim, Daryl Chrzan, Aron Walsh, Samuel~M. Blau, Mark Asta, and Aditi~S. Krishnapriyan.
\newblock {{MLIP Arena}}: {{Advancing Fairness}} and {{Transparency}} in {{Machine Learning Interatomic Potentials}} via an {{Open}}, {{Accessible Benchmark Platform}}.
\newblock In \emph{The {{Thirty-ninth Annual Conference}} on {{Neural Information Processing Systems Datasets}} and {{Benchmarks Track}}}, 2025.
\newblock URL \url{https://openreview.net/forum?id=SAT0KPA5UO}.

\bibitem[P\'{o}ta et~al.(2025)P\'{o}ta, Ahlawat, Cs\'{a}nyi, and Simoncelli]{Póta-2025}
Bal\'{a}zs P\'{o}ta, Paramvir Ahlawat, G\'{a}bor Cs\'{a}nyi, and Michele Simoncelli.
\newblock Thermal {{Conductivity Predictions}} with {{Foundation Atomistic Models}}, 2025.
\newblock URL \url{http://arxiv.org/abs/2408.00755}.

\bibitem[Yu et~al.(2024)Yu, Giantomassi, Materzanini, Wang, and Rignanese]{Yu-2024}
Haochen Yu, Matteo Giantomassi, Giuliana Materzanini, Junjie Wang, and Gian-Marco Rignanese.
\newblock Systematic assessment of various universal machine-learning interatomic potentials.
\newblock \emph{Materials Genome Engineering Advances}, 2\penalty0 (3):\penalty0 e58, 2024.
\newblock ISSN 2940-9497.
\newblock \doi{10.1002/mgea.58}.
\newblock URL \url{https://onlinelibrary.wiley.com/doi/abs/10.1002/mgea.58}.

\bibitem[Fu et~al.(2025)Fu, Wood, Barroso-Luque, Levine, Gao, Dzamba, and Zitnick]{Fu-2025}
Xiang Fu, Brandon~M. Wood, Luis Barroso-Luque, Daniel~S. Levine, Meng Gao, Misko Dzamba, and C.~Lawrence Zitnick.
\newblock Learning {{Smooth}} and {{Expressive Interatomic Potentials}} for {{Physical Property Prediction}}.
\newblock In \emph{Proceedings of the 42nd {{International Conference}} on {{Machine Learning}}}, pages 17875--17893. PMLR, 2025.
\newblock URL \url{https://proceedings.mlr.press/v267/fu25h.html}.

\bibitem[Fu et~al.(2023)Fu, Wu, Wang, Xie, Keten, Gomez-Bombarelli, and Jaakkola]{Fu-2023}
Xiang Fu, Zhenghao Wu, Wujie Wang, Tian Xie, Sinan Keten, Rafael Gomez-Bombarelli, and Tommi Jaakkola.
\newblock Forces are not {{Enough}}: {{Benchmark}} and {{Critical Evaluation}} for {{Machine Learning Force Fields}} with {{Molecular Simulations}}.
\newblock \emph{Transactions on Machine Learning Research}, 2023.
\newblock ISSN 2835-8856.
\newblock URL \url{https://openreview.net/forum?id=A8pqQipwkt}.

\bibitem[Miksch et~al.(2021-07)Miksch, Morawietz, Kästner, Urban, and Artrith]{Miksch-2021}
April~M Miksch, Tobias Morawietz, Johannes Kästner, Alexander Urban, and Nongnuch Artrith.
\newblock Strategies for the construction of machine-learning potentials for accurate and efficient atomic-scale simulations.
\newblock \emph{Machine Learning: Science and Technology}, 2\penalty0 (3):\penalty0 031001, 2021-07.
\newblock ISSN 2632-2153.
\newblock \doi{10.1088/2632-2153/abfd96}.
\newblock URL \url{https://doi.org/10.1088/2632-2153/abfd96}.

\bibitem[Stocker et~al.(2022-11)Stocker, Gasteiger, Becker, G\"{u}nnemann, and Margraf]{Stocker-2022}
Sina Stocker, Johannes Gasteiger, Florian Becker, Stephan G\"{u}nnemann, and Johannes~T Margraf.
\newblock How robust are modern graph neural network potentials in long and hot molecular dynamics simulations?
\newblock \emph{Machine Learning: Science and Technology}, 3\penalty0 (4):\penalty0 045010, 2022-11.
\newblock ISSN 2632-2153.
\newblock \doi{10.1088/2632-2153/ac9955}.
\newblock URL \url{https://doi.org/10.1088/2632-2153/ac9955}.

\bibitem[Liu et~al.(2023)Liu, He, and Mo]{Liu-2023}
Yunsheng Liu, Xingfeng He, and Yifei Mo.
\newblock Discrepancies and error evaluation metrics for machine learning interatomic potentials.
\newblock \emph{npj Computational Materials}, 9\penalty0 (1):\penalty0 174, 2023.
\newblock ISSN 2057-3960.
\newblock \doi{10.1038/s41524-023-01123-3}.
\newblock URL \url{https://www.nature.com/articles/s41524-023-01123-3}.

\bibitem[Liu and Mo(2024)]{Liu-2024}
Yunsheng Liu and Yifei Mo.
\newblock Learning from models: High-dimensional analyses on the performance of machine learning interatomic potentials.
\newblock \emph{npj Computational Materials}, 10\penalty0 (1):\penalty0 159, 2024.
\newblock ISSN 2057-3960.
\newblock \doi{10.1038/s41524-024-01333-3}.
\newblock URL \url{https://www.nature.com/articles/s41524-024-01333-3}.

\bibitem[Witt et~al.(2023)Witt, van~der Oord, Gel\v{z}inyt\.{e}, J\"{a}rvinen, Ross, Darby, Ho, Baldwin, Sachs, Kermode, Bernstein, Cs\'{a}nyi, and Ortner]{Witt-2023}
William~C. Witt, Cas van~der Oord, Elena Gel\v{z}inyt\.{e}, Teemu J\"{a}rvinen, Andres Ross, James~P. Darby, Cheuk~Hin Ho, William~J. Baldwin, Matthias Sachs, James Kermode, Noam Bernstein, G\'{a}bor Cs\'{a}nyi, and Christoph Ortner.
\newblock {{ACEpotentials}}.jl: {{A Julia}} implementation of the atomic cluster expansion.
\newblock \emph{The Journal of Chemical Physics}, 159\penalty0 (16):\penalty0 164101, 2023.
\newblock ISSN 0021-9606.
\newblock \doi{10.1063/5.0158783}.
\newblock URL \url{https://doi.org/10.1063/5.0158783}.

\bibitem[Rhodes et~al.(2025)Rhodes, Vandenhaute, \v{S}imkus, Gin, Godwin, Duignan, and Neumann]{Rhodes-2025}
Benjamin Rhodes, Sander Vandenhaute, Vaidotas \v{S}imkus, James Gin, Jonathan Godwin, Tim Duignan, and Mark Neumann.
\newblock Orb-v3: Atomistic simulation at scale, 2025.
\newblock URL \url{http://arxiv.org/abs/2504.06231}.

\bibitem[Morrow et~al.(2023)Morrow, Gardner, and Deringer]{Morrow-2023}
Joe~D. Morrow, John L.~A. Gardner, and Volker~L. Deringer.
\newblock How to validate machine-learned interatomic potentials.
\newblock \emph{The Journal of Chemical Physics}, 158\penalty0 (12):\penalty0 121501, 2023.
\newblock ISSN 0021-9606.
\newblock \doi{10.1063/5.0139611}.
\newblock URL \url{https://doi.org/10.1063/5.0139611}.

\bibitem[Deng et~al.(2025)Deng, Choi, Zhong, Riebesell, Anand, Li, Jun, Persson, and Ceder]{Deng-2025}
Bowen Deng, Yunyeong Choi, Peichen Zhong, Janosh Riebesell, Shashwat Anand, Zhuohan Li, KyuJung Jun, Kristin~A. Persson, and Gerbrand Ceder.
\newblock Systematic softening in universal machine learning interatomic potentials.
\newblock \emph{npj Computational Materials}, 11\penalty0 (1):\penalty0 9, 2025.
\newblock ISSN 2057-3960.
\newblock \doi{10.1038/s41524-024-01500-6}.
\newblock URL \url{https://www.nature.com/articles/s41524-024-01500-6}.

\bibitem[Cărare et~al.(2025)Cărare, Thiemann, Morrow, Wales, Pyzer-Knapp, and Dicks]{Cărare-2025}
Vlad Cărare, Fabian~L. Thiemann, Joe~D. Morrow, David~J. Wales, Edward~O. Pyzer-Knapp, and Luke Dicks.
\newblock Global properties of the energy landscape: A testing and training arena for machine learned potentials.
\newblock \emph{npj Computational Materials}, 12\penalty0 (1):\penalty0 9, 2025.
\newblock ISSN 2057-3960.
\newblock \doi{10.1038/s41524-025-01878-x}.
\newblock URL \url{https://www.nature.com/articles/s41524-025-01878-x}.

\bibitem[Ranasinghe et~al.(2025)Ranasinghe, Baskerville, Wood, and K\"{o}nig]{Ranasinghe-2025}
Kavindri Ranasinghe, Adam~L. Baskerville, Geoffrey P.~F. Wood, and Gerhard K\"{o}nig.
\newblock Basic {{Stability Tests}} of {{Machine Learning Potentials}} for {{Molecular Simulations}} in {{Computational Drug Discovery}}.
\newblock \emph{Journal of Chemical Information and Modeling}, 65\penalty0 (17):\penalty0 8980--8999, 2025.
\newblock ISSN 1549-9596.
\newblock \doi{10.1021/acs.jcim.5c01150}.
\newblock URL \url{https://doi.org/10.1021/acs.jcim.5c01150}.

\bibitem[Thomas et~al.(2018)Thomas, Smidt, Kearnes, Yang, Li, Kohlhoff, and Riley]{Thomas-2018}
Nathaniel Thomas, Tess Smidt, Steven Kearnes, Lusann Yang, Li~Li, Kai Kohlhoff, and Patrick Riley.
\newblock Tensor field networks: {{Rotation-}} and translation-equivariant neural networks for {{3D}} point clouds, 2018.
\newblock URL \url{http://arxiv.org/abs/1802.08219}.

\bibitem[Geiger and Smidt(2022)]{Geiger-2022}
Mario Geiger and Tess Smidt.
\newblock E3nn: {{Euclidean Neural Networks}}, 2022.
\newblock URL \url{http://arxiv.org/abs/2207.09453}.

\bibitem[Frank et~al.(2024)Frank, Unke, M\"{u}ller, and Chmiela]{Frank-2024}
J.~Thorben Frank, Oliver~T. Unke, Klaus-Robert M\"{u}ller, and Stefan Chmiela.
\newblock A {{Euclidean}} transformer for fast and stable machine learned force fields.
\newblock \emph{Nature Communications}, 15\penalty0 (1):\penalty0 6539, 2024.
\newblock ISSN 2041-1723.
\newblock \doi{10.1038/s41467-024-50620-6}.
\newblock URL \url{https://www.nature.com/articles/s41467-024-50620-6}.

\bibitem[Xie et~al.(2025)Xie, Daigavane, Kotak, and Smidt]{Xie-2025}
YuQing Xie, Ameya Daigavane, Mit Kotak, and Tess Smidt.
\newblock The {{Price}} of {{Freedom}}: {{Exploring Expressivity}} and {{Runtime Tradeoffs}} in {{Equivariant Tensor Products}}.
\newblock In \emph{Forty-Second {{International Conference}} on {{Machine Learning}}}, 2025.
\newblock URL \url{https://openreview.net/forum?id=EvIwwGYTLc&referrer=%5Bthe%20profile%20of%20Tess%20Smidt%5D(%2Fprofile%3Fid%3D~Tess_Smidt1)}.

\bibitem[Liao et~al.(2023{\natexlab{a}})Liao, Smidt, and Das]{Liao-2023a}
Yi-Lun Liao, Tess Smidt, and Abhishek Das.
\newblock Generalizing {{Denoising}} to {{Non-Equilibrium Structures Improves Equivariant Force Fields}}.
\newblock 2023{\natexlab{a}}.
\newblock URL \url{https://openreview.net/forum?id=X7gqOBG8ow}.

\bibitem[Liao et~al.(2023{\natexlab{b}})Liao, Wood, Das, and Smidt]{Liao-2023}
Yi-Lun Liao, Brandon~M. Wood, Abhishek Das, and Tess Smidt.
\newblock {{EquiformerV2}}: {{Improved Equivariant Transformer}} for {{Scaling}} to {{Higher-Degree Representations}}.
\newblock In \emph{The {{Twelfth International Conference}} on {{Learning Representations}}}, 2023{\natexlab{b}}.
\newblock URL \url{https://openreview.net/forum?id=mCOBKZmrzD}.

\bibitem[Cen et~al.(2025)Cen, Li, Lin, Ren, Wang, and Huang]{Cen-2025}
Jiacheng Cen, Anyi Li, Ning Lin, Yuxiang Ren, Zihe Wang, and Wenbing Huang.
\newblock Are {{High-Degree Representations Really Unnecessary}} in {{Equivariant Graph Neural Networks}}?, 2025.
\newblock URL \url{http://arxiv.org/abs/2410.11443}.

\bibitem[Geiger et~al.(2024-11-18T18:30:00+00:00)Geiger, Kucukbenli, Zandstein, and Tretina]{Geiger-2024}
Mario Geiger, Emine Kucukbenli, Becca Zandstein, and Kyle Tretina.
\newblock Accelerate {{Drug}} and {{Material Discovery}} with {{New Math Library NVIDIA cuEquivariance}}, 2024-11-18T18:30:00+00:00.
\newblock URL \url{https://developer.nvidia.com/blog/accelerate-drug-and-material-discovery-with-new-math-library-nvidia-cuequivariance/}.

\bibitem[Tan et~al.(2025)Tan, Descoteaux, Kotak, Nascimento, Kavanagh, Zichi, Wang, Saluja, Hu, Smidt, Johansson, Witt, Kozinsky, and Musaelian]{Tan-2025}
Chuin~Wei Tan, Marc~L. Descoteaux, Mit Kotak, Gabriel de~Miranda Nascimento, Se\'{a}n~R. Kavanagh, Laura Zichi, Menghang Wang, Aadit Saluja, Yizhong~R. Hu, Tess Smidt, Anders Johansson, William~C. Witt, Boris Kozinsky, and Albert Musaelian.
\newblock High-performance training and inference for deep equivariant interatomic potentials, 2025.
\newblock URL \url{http://arxiv.org/abs/2504.16068}.

\bibitem[Lee et~al.(2025)Lee, Kim, Park, Jeong, Han, Park, and Lee]{Lee-2025}
Seung~Yul Lee, Hojoon Kim, Yutack Park, Dawoon Jeong, Seungwu Han, Yeonhong Park, and Jae~W. Lee.
\newblock {{FlashTP}}: {{Fused}}, {{Sparsity-Aware Tensor Product}} for {{Machine Learning Interatomic Potentials}}.
\newblock In \emph{Forty-Second {{International Conference}} on {{Machine Learning}}}, 2025.
\newblock URL \url{https://openreview.net/forum?id=wiQe95BPaB}.

\bibitem[Passaro and Zitnick(2023)]{Passaro-2023}
Saro Passaro and C.~Lawrence Zitnick.
\newblock Reducing {{SO}}(3) convolutions to {{SO}}(2) for efficient equivariant {{GNNs}}.
\newblock In \emph{Proceedings of the 40th {{International Conference}} on {{Machine Learning}}}, volume 202 of \emph{{{ICML}}'23}, pages 27420--27438, Honolulu, Hawaii, USA, 2023. JMLR.org.

\bibitem[Frank et~al.(2022)Frank, Unke, and Müller]{Frank-2022}
J.~Thorben Frank, Oliver~T. Unke, and Klaus-Robert Müller.
\newblock So3krates: Equivariant attention for interactions on arbitrary length-scales in molecular systems.
\newblock In \emph{Proceedings of the 36th {{International Conference}} on {{Neural Information Processing Systems}}}, {{NIPS}} '22, pages 29400--29413, Red Hook, NY, USA, 2022. Curran Associates Inc.
\newblock ISBN 978-1-7138-7108-8.

\bibitem[Bharadwaj et~al.(2025)Bharadwaj, Glover, Buluc, and Demmel]{Bharadwaj-2025}
Vivek Bharadwaj, Austin Glover, Aydin Buluc, and James Demmel.
\newblock An {{Efficient Sparse Kernel Generator}} for {{O}}(3)-{{Equivariant Deep Networks}}, 2025.
\newblock URL \url{http://arxiv.org/abs/2501.13986}.

\bibitem[Aykent and Xia(2024)]{Aykent-2024}
Sarp Aykent and Tian Xia.
\newblock {{GotenNet}}: {{Rethinking Efficient 3D Equivariant Graph Neural Networks}}.
\newblock In \emph{The {{Thirteenth International Conference}} on {{Learning Representations}}}, 2024.
\newblock URL \url{https://openreview.net/forum?id=5wxCQDtbMo}.

\bibitem[Zhang and Guo(2025)]{Zhang-2025}
Yaolong Zhang and Hua Guo.
\newblock Node-equivariant message passing for efficient and accurate machine learning interatomic potentials.
\newblock \emph{Chemical Science}, 2025.
\newblock ISSN 2041-6539.
\newblock \doi{10.1039/D5SC07248D}.
\newblock URL \url{https://pubs.rsc.org/en/content/articlelanding/2026/sc/d5sc07248d}.

\bibitem[Chmiela et~al.(2022)Chmiela, Vassilev-Galindo, Unke, Kabylda, Sauceda, Tkatchenko, and M\"{u}ller]{Chmiela-2022}
Stefan Chmiela, Valentin Vassilev-Galindo, Oliver~T. Unke, Adil Kabylda, Huziel~E. Sauceda, Alexandre Tkatchenko, and Klaus-Robert M\"{u}ller.
\newblock Accurate global machine learning force fields for molecules with hundreds of atoms, 2022.
\newblock URL \url{http://arxiv.org/abs/2209.14865}.

\bibitem[Christensen and von Lilienfeld(2020)]{Christensen-2020}
Anders~S. Christensen and O.~Anatole von Lilienfeld.
\newblock On the role of gradients for machine learning of molecular energies and forces, 2020.
\newblock URL \url{http://arxiv.org/abs/2007.09593}.

\bibitem[Horton et~al.(2025)Horton, Huck, Yang, Munro, Dwaraknath, Ganose, Kingsbury, Wen, Shen, Mathis, et~al.]{Horton-2025}
Matthew~K Horton, Patrick Huck, Ruo~Xi Yang, Jason~M Munro, Shyam Dwaraknath, Alex~M Ganose, Ryan~S Kingsbury, Mingjian Wen, Jimmy~X Shen, Tyler~S Mathis, et~al.
\newblock Accelerated data-driven materials science with the materials project.
\newblock \emph{Nature Materials}, pages 1--11, 2025.

\bibitem[Loew et~al.(2025)Loew, Sun, Wang, Botti, and Marques]{Loew-2025}
Antoine Loew, Dewen Sun, Hai-Chen Wang, Silvana Botti, and Miguel A.~L. Marques.
\newblock Universal machine learning interatomic potentials are ready for phonons.
\newblock \emph{npj Computational Materials}, 11\penalty0 (1):\penalty0 178, 2025.
\newblock ISSN 2057-3960.
\newblock \doi{10.1038/s41524-025-01650-1}.
\newblock URL \url{https://www.nature.com/articles/s41524-025-01650-1}.

\bibitem[L\'{o}pez et~al.(2024)L\'{o}pez, Rurali, and Cazorla]{López-2024}
Cibr\'{a}n L\'{o}pez, Riccardo Rurali, and Claudio Cazorla.
\newblock How {{Concerted Are Ionic Hops}} in {{Inorganic Solid-State Electrolytes}}?
\newblock \emph{Journal of the American Chemical Society}, 146\penalty0 (12):\penalty0 8269--8279, 2024.
\newblock ISSN 0002-7863.
\newblock \doi{10.1021/jacs.3c13279}.
\newblock URL \url{https://doi.org/10.1021/jacs.3c13279}.

\bibitem[Gasteiger et~al.(2019)Gasteiger, Groß, and Günnemann]{Gasteiger-2019}
Johannes Gasteiger, Janek Groß, and Stephan Günnemann.
\newblock Directional {{Message Passing}} for {{Molecular Graphs}}.
\newblock In \emph{International {{Conference}} on {{Learning Representations}}}, 2019.
\newblock URL \url{https://openreview.net/forum?id=B1eWbxStPH}.

\bibitem[Batatia et~al.(2025-01)Batatia, Batzner, Kov\'{a}cs, Musaelian, Simm, Drautz, Ortner, Kozinsky, and Cs\'{a}nyi]{Batatia-2025a}
Ilyes Batatia, Simon Batzner, D\'{a}vid~P\'{e}ter Kov\'{a}cs, Albert Musaelian, Gregor N.~C. Simm, Ralf Drautz, Christoph Ortner, Boris Kozinsky, and G\'{a}bor Cs\'{a}nyi.
\newblock The design space of {{E}}(3)-equivariant atom-centred interatomic potentials.
\newblock \emph{Nature Machine Intelligence}, 7\penalty0 (1):\penalty0 56--67, 2025-01.
\newblock ISSN 2522-5839.
\newblock \doi{10.1038/s42256-024-00956-x}.
\newblock URL \url{https://www.nature.com/articles/s42256-024-00956-x}.

\bibitem[Thölke and Fabritiis(2021)]{Thölke-2021}
Philipp Thölke and Gianni~De Fabritiis.
\newblock Equivariant {{Transformers}} for {{Neural Network}} based {{Molecular Potentials}}.
\newblock In \emph{International {{Conference}} on {{Learning Representations}}}, 2021.
\newblock URL \url{https://openreview.net/forum?id=zNHzqZ9wrRB}.

\bibitem[Ziegler and Biersack(1985)]{Ziegler-1985}
James~F. Ziegler and Jochen~P. Biersack.
\newblock The {{Stopping}} and {{Range}} of {{Ions}} in {{Matter}}.
\newblock In D.~Allan Bromley, editor, \emph{Treatise on {{Heavy-Ion Science}}}, pages 93--129. Springer US, Boston, MA, 1985.
\newblock ISBN 978-1-4615-8105-5.
\newblock \doi{10.1007/978-1-4615-8103-1_3}.

\bibitem[Ghahremanpour et~al.(2018)Ghahremanpour, van Maaren, and van~der Spoel]{Ghahremanpour-2018}
Mohammad~M. Ghahremanpour, Paul~J. van Maaren, and David van~der Spoel.
\newblock The {{Alexandria}} library, a quantum-chemical database of molecular properties for force field development.
\newblock \emph{Scientific Data}, 5\penalty0 (1):\penalty0 180062, 2018.
\newblock ISSN 2052-4463.
\newblock \doi{10.1038/sdata.2018.62}.
\newblock URL \url{https://www.nature.com/articles/sdata201862}.

\bibitem[Bochkarev et~al.(2024)Bochkarev, Lysogorskiy, and Drautz]{Bochkarev-2024}
Anton Bochkarev, Yury Lysogorskiy, and Ralf Drautz.
\newblock Graph {{Atomic Cluster Expansion}} for {{Semilocal Interactions}} beyond {{Equivariant Message Passing}}.
\newblock \emph{Physical Review X}, 14\penalty0 (2):\penalty0 021036, 2024.
\newblock \doi{10.1103/PhysRevX.14.021036}.
\newblock URL \url{https://link.aps.org/doi/10.1103/PhysRevX.14.021036}.

\bibitem[Cohen et~al.(2025-11)Cohen, Riebesell, Goodall, Kolluru, Falletta, Krause, Colindres, Ceder, and Gangan]{Cohen-2025}
Orion Cohen, Janosh Riebesell, Rhys Goodall, Adeesh Kolluru, Stefano Falletta, Joseph Krause, Jorge Colindres, Gerbrand Ceder, and Abhijeet~S Gangan.
\newblock {{TorchSim}}: An efficient atomistic simulation engine in {{PyTorch}}.
\newblock \emph{AI for Science}, 1\penalty0 (2):\penalty0 025003, 2025-11.
\newblock ISSN 3050-287X.
\newblock \doi{10.1088/3050-287X/ae1799}.
\newblock URL \url{https://doi.org/10.1088/3050-287X/ae1799}.

\bibitem[NVI(2025-10-20T16:30:00+00:00)]{NVIDIA-2025}
Enabling {{Scalable AI-Driven Molecular Dynamics Simulations}}, 2025-10-20T16:30:00+00:00.
\newblock URL \url{https://developer.nvidia.com/blog/enabling-scalable-ai-driven-molecular-dynamics-simulations/}.

\bibitem[Famprikis et~al.(2019)Famprikis, Canepa, Dawson, Islam, and Masquelier]{Famprikis-2019}
Theodosios Famprikis, Pieremanuele Canepa, James~A. Dawson, M.~Saiful Islam, and Christian Masquelier.
\newblock Fundamentals of inorganic solid-state electrolytes for batteries.
\newblock \emph{Nature Materials}, 18\penalty0 (12):\penalty0 1278--1291, December 2019.
\newblock ISSN 1476-4660.
\newblock \doi{10.1038/s41563-019-0431-3}.

\bibitem[Schran et~al.(2021)Schran, Thiemann, Rowe, M\"{u}ller, Marsalek, and Michaelides]{Schran-2021}
Christoph Schran, Fabian~L. Thiemann, Patrick Rowe, Erich~A. M\"{u}ller, Ondrej Marsalek, and Angelos Michaelides.
\newblock Machine learning potentials for complex aqueous systems made simple.
\newblock \emph{Proceedings of the National Academy of Sciences}, 118\penalty0 (38):\penalty0 e2110077118, 2021.
\newblock \doi{10.1073/pnas.2110077118}.
\newblock URL \url{https://www.pnas.org/doi/abs/10.1073/pnas.2110077118}.

\bibitem[Fragapane and Deringer(2025)]{Fragapane-2025}
Natascia~L. Fragapane and Volker~L. Deringer.
\newblock Li-{{P-S Electrolyte Materials}} as a {{Benchmark}} for {{Machine-Learned Interatomic Potentials}}, 2025.
\newblock URL \url{http://arxiv.org/abs/2511.16569}.

\bibitem[Kingma and Ba(2017)]{Kingma-2017}
Diederik~P. Kingma and Jimmy Ba.
\newblock Adam: {{A Method}} for {{Stochastic Optimization}}, 2017.
\newblock URL \url{http://arxiv.org/abs/1412.6980}.

\bibitem[Loshchilov and Hutter(2019)]{Loshchilov-2019}
Ilya Loshchilov and Frank Hutter.
\newblock Decoupled {{Weight Decay Regularization}}, 2019.
\newblock URL \url{http://arxiv.org/abs/1711.05101}.

\bibitem[Kumar et~al.(2025)Kumar, Owen, Chowdhury, and G\"{u}ra]{Kumar-2025}
Abhay Kumar, Louis Owen, Nilabhra~Roy Chowdhury, and Fabian G\"{u}ra.
\newblock {{ZClip}}: {{Adaptive Spike Mitigation}} for {{LLM Pre-Training}}, 2025.
\newblock URL \url{http://arxiv.org/abs/2504.02507}.

\bibitem[Hjorth~Larsen et~al.(2017-06)Hjorth~Larsen, J\o{}rgen~Mortensen, Blomqvist, Castelli, Christensen, Du\l{}ak, Friis, Groves, Hammer, Hargus, Hermes, Jennings, Bjerre~Jensen, Kermode, Kitchin, Leonhard~Kolsbjerg, Kubal, Kaasbjerg, Lysgaard, Bergmann~Maronsson, Maxson, Olsen, Pastewka, Peterson, Rostgaard, Schi\o{}tz, Sch\"{u}tt, Strange, Thygesen, Vegge, Vilhelmsen, Walter, Zeng, and Jacobsen]{HjorthLarsen-2017}
Ask Hjorth~Larsen, Jens J\o{}rgen~Mortensen, Jakob Blomqvist, Ivano~E Castelli, Rune Christensen, Marcin Du\l{}ak, Jesper Friis, Michael~N Groves, Bj\o{}rk Hammer, Cory Hargus, Eric~D Hermes, Paul~C Jennings, Peter Bjerre~Jensen, James Kermode, John~R Kitchin, Esben Leonhard~Kolsbjerg, Joseph Kubal, Kristen Kaasbjerg, Steen Lysgaard, J\'{o}n Bergmann~Maronsson, Tristan Maxson, Thomas Olsen, Lars Pastewka, Andrew Peterson, Carsten Rostgaard, Jakob Schi\o{}tz, Ole Sch\"{u}tt, Mikkel Strange, Kristian~S Thygesen, Tejs Vegge, Lasse Vilhelmsen, Michael Walter, Zhenhua Zeng, and Karsten~W Jacobsen.
\newblock The atomic simulation environment--a {{Python}} library for working with atoms.
\newblock \emph{Journal of Physics: Condensed Matter}, 29\penalty0 (27):\penalty0 273002, 2017-06.
\newblock ISSN 0953-8984.
\newblock \doi{10.1088/1361-648X/aa680e}.
\newblock URL \url{https://doi.org/10.1088/1361-648X/aa680e}.

\bibitem[Perdew et~al.(1996)Perdew, Burke, and Ernzerhof]{Perdew-1996}
John~P. Perdew, Kieron Burke, and Matthias Ernzerhof.
\newblock Generalized {{Gradient Approximation Made Simple}}.
\newblock \emph{Physical Review Letters}, 77\penalty0 (18):\penalty0 3865--3868, 1996.
\newblock \doi{10.1103/PhysRevLett.77.3865}.
\newblock URL \url{https://link.aps.org/doi/10.1103/PhysRevLett.77.3865}.

\bibitem[Nos\'{e}(1984)]{Nosé-1984}
Shuichi Nos\'{e}.
\newblock A unified formulation of the constant temperature molecular dynamics methods.
\newblock \emph{The Journal of Chemical Physics}, 81\penalty0 (1):\penalty0 511--519, 1984.
\newblock ISSN 0021-9606.
\newblock \doi{10.1063/1.447334}.
\newblock URL \url{https://doi.org/10.1063/1.447334}.

\end{thebibliography}





\makeatletter

\newcommand{\makesititle}{%
  \vbox{%
    \hsize\textwidth
    \linewidth\hsize
    \vskip 0.1in
    {\LARGE \@title\par}
    \vskip 0.1in
    \vskip 0.2in
  }
}
\makeatother

\title{Supplementary Information
}

\author{}
\renewcommand{\undertitle}{}
\renewcommand{\headeright}{}

\makesititle

\setcounter{page}{1}
\setcounter{equation}{0}
\setcounter{figure}{0}
\renewcommand{\thepage}{S\arabic{page}} 
\renewcommand{\thetable}{S\arabic{table}}  
\renewcommand{\thefigure}{S\arabic{figure}}
\renewcommand{\thesection}{\Alph{section}}
\renewcommand{\thesubsection}{\thesection.\arabic{subsection}}

In this supplementary information, we provide further details on specific aspects of the study presented in the manuscript. This includes architectural design considerations for attention mechanisms and smoothness, evaluation of diatomic binding curves across the periodic table, sAlex benchmark, and characterization of solid-state electrolyte systems.

\tableofcontents

\section{Diatomic Binding Curves}

\begin{figure}[h]
    \centering
    \includegraphics[width=1\linewidth]{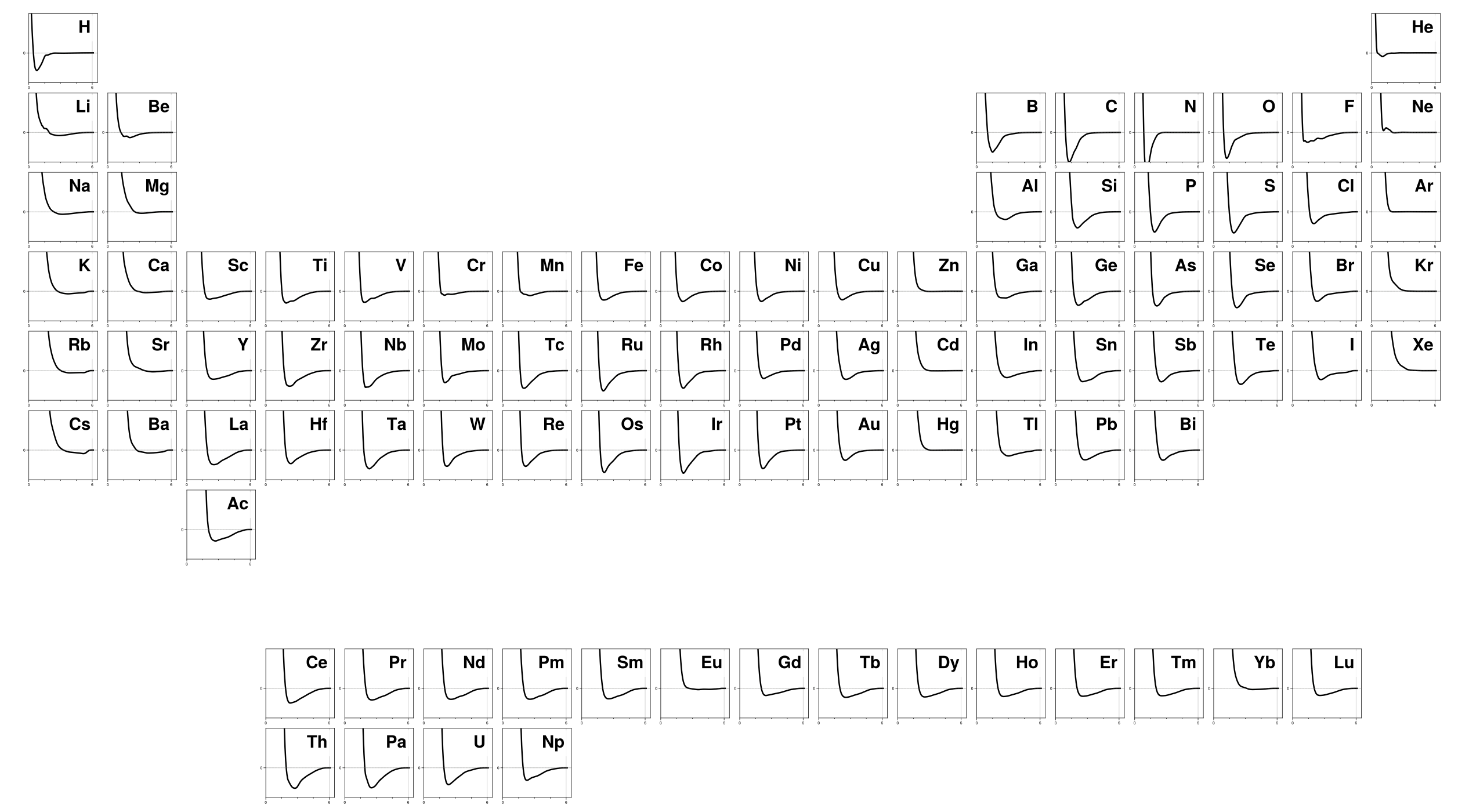}
    \caption{\textbf{Diatomics binding curves across elements in MPtrj.}}
    \label{fig:diatomics-pt}
\end{figure}

Figure~\ref{fig:diatomics-pt} presents the complete set of homonuclear diatomic binding curves for all elements in the MPtrj dataset. The main text analyzes representative cases (H, P, Cu) and quantitative smoothness metrics. These comprehensive curves demonstrate that Geodite-MP produces physically consistent short-range repulsion and smooth potential energy surfaces across the full range of chemical elements encountered during training, with no explicit diatomic configurations in the training data.

\section{sAlex}

The main manuscript evaluates Geodite-MP primarily on materials from the MPtrj training distribution. To assess out-of-equilibrium generalization, we evaluated multiple models on a subset of 10,000 randomly selected systems from the sAlex benchmark, which comprises diverse inorganic materials not captured in the MPtrj dataset. This benchmark tests the ability to predict a system energy, forces and stress in an unbiased fashion, due to the lack of test set from the MPtrj dataset. Geodite-MP achieves an energy MAE of 35.09 meV/atom, force MAE of 90.33 meV/Å, and stress MAE of 0.38 meV/Å$^3$/atom. These results place Geodite-MP between the top-performing models (Allegro-MP-L and eSEN-30M-MP at approximately 34 meV/atom) and mid-tier architectures, demonstrating generalization to materials outside its training distribution.

\begin{table}[H]
    \caption{\textbf{Out-of-distribution generalization on sAlex benchmark.}
Mean absolute errors for energy, forces, and stress on 10,000 randomly selected inorganic structures from sAlex dataset. Systems not present in MPtrj training distribution. All models trained on MPtrj. Lower values indicate better generalization to unseen materials.}
    \centering
    \resizebox{\textwidth*2/3}{!}{
        \begin{tabular}{lccc}
        \toprule
        \midrule
        Model & \shortstack{Energy MAE $\downarrow$\\ \footnotesize (meV/atom)} & \shortstack{Force MAE $\downarrow$\\ \footnotesize (meV/\AA)} & \shortstack{Stress MAE $\downarrow$\\ \footnotesize (meV/\AA$^3$/atom)} \\
        \midrule
        \midrule
        Allegro-MP-L & 33.79 & 72.51 & 0.33 \\
        eSEN-30M-MP & 33.88 & 73.12 & 0.37 \\
        Eqnorm MPtrj & 35.08 & 82.05 & 0.35 \\
        Nequip-MP-L & 37.25 & 85.97 & 0.39 \\
        SevenNet-l3i5 & 41.09 & 93.06 & 0.42 \\
        MACE-MP-0 & 53.25 & 85.64 & 0.45 \\
        GRACE-2L-MPtrj & 80.90 & 89.70 & 0.48 \\
        \midrule
        Geodite-MP & 35.09 & 90.33 & 0.38 \\
        \midrule
        \bottomrule
        \end{tabular}
    }
\end{table}

\section{SSE Systems}

Table~\ref{tab:material_summary} provides an overview of the 49 solid-state electrolyte simulations used to assess long-term molecular dynamics stability. The systems span a range of compositions, including lithium-, sodium-, and copper-based ionic conductors, with simulation temperatures varying from 300 to 1300~K and system sizes between 128 and 384 atoms.

\begin{longtable}{lrrrr}
\caption{Summary of materials, their number of atoms, trajectory details, and RDF scores for various models.}
\label{tab:material_summary} \\
\toprule
Material & Number of Atoms & Temperature (K) & Frames & Trajectory Length (ps) \\
\midrule
\endfirsthead

\caption[]{Summary of materials, their number of atoms, trajectory details, and RDF scores for various models.} \\
\toprule
Material & Number of Atoms & Temperature (K) & Frames & Trajectory Length (ps) \\
\midrule
\endhead

\midrule
\multicolumn{5}{r}{Continued on next page} \\
\midrule
\endfoot

\bottomrule
\endlastfoot
CsPbBr3 & 180 & 500 & 4031 & 60.465 \\
CsPbBr3 & 180 & 700 & 4895 & 73.425 \\
CuI     & 216 & 500 & 4087 & 61.305 \\
CuI     & 216 & 700 & 4056 & 60.840 \\
CuI     & 216 & 900 & 4015 & 60.225 \\
Li10GeS2P12  & 200 & 650 & 9966 & 149.490 \\
Li10GeS2P12  & 200 & 900 & 9840 & 147.600 \\
Li10GeS2P12  & 200 & 1150 & 9737 & 146.055 \\
Li2SnS3      & 384 & 400 & 5346 & 80.190 \\
Li2SnS3      & 384 & 600 & 5303 & 79.545 \\
Li2SnS3      & 384 & 800 & 5199 & 77.985 \\
Li3La4Ti8O24 & 312 & 500 & 6900 & 103.500 \\
Li3La4Ti8O24 & 312 & 700 & 7812 & 117.180 \\
Li3La4Ti8O24 & 312 & 900 & 7511 & 112.665 \\
Li3La4Ti8O24 & 312 & 1100 & 7044 & 105.660 \\
Li3N & 256 & 300 & 7024 & 105.360 \\
Li3N & 256 & 500 & 6228 & 93.420 \\
Li3N & 256 & 700 & 3628 & 54.420 \\
Li3N & 256 & 900 & 3030 & 45.450 \\
Li3N & 256 & 1100 & 2845 & 42.675 \\
Li3OCl & 320 & 1250 & 7595 & 113.925 \\
Li7P3S11 & 336 & 400 & 6959 & 104.385 \\
Li7P3S11 & 336 & 600 & 6616 & 99.240 \\
Li7P3S11 & 336 & 800 & 6250 & 93.750 \\
LiGaO2 & 128 & 300 & 8383 & 125.745 \\
LiGaO2 & 128 & 500 & 8038 & 120.570 \\
LiGaO2 & 128 & 700 & 12353 & 185.295 \\
LiGaO2 & 128 & 900 & 11853 & 177.795 \\
LiGaO2 & 128 & 1100 & 14471 & 217.065 \\
LiGaO2 & 128 & 1300 & 11052 & 165.780 \\
LiIO3 & 270 & 700 & 5827 & 87.405 \\
LiIO3 & 270 & 900 & 4894 & 73.410 \\
LiIO3 & 270 & 1100 & 6574 & 98.610 \\
LiNbO3 & 180 & 500 & 11957 & 179.355 \\
LiNbO3 & 180 & 700 & 7664 & 114.960 \\
LiNbO3 & 180 & 900 & 7200 & 108.000 \\
Na2B4O7 & 312 & 300 & 8145 & 122.175 \\
Na2B4O7 & 312 & 450 & 7874 & 118.110 \\
Na2B4O7 & 312 & 600 & 8328 & 124.920 \\
Na2B4O7 & 312 & 750 & 6754 & 101.310 \\
Na3SbS4 & 288 & 500 & 5372 & 80.580 \\
Na3SbS4 & 288 & 700 & 5454 & 81.810 \\
Na3SbS4 & 288 & 900 & 5081 & 76.215 \\
NaBH4 & 288 & 300 & 12801 & 192.015 \\
NaBH4 & 288 & 450 & 12757 & 191.355 \\
NaBH4 & 288 & 600 & 11971 & 179.565 \\
NaBH4 & 288 & 750 & 11776 & 176.640 \\
NaBO2 & 288 & 750 & 3540 & 53.100 \\
d-Bi2O3 & 180 & 500 & 13214 & 198.210 \\
\end{longtable}

\end{document}